\documentclass[11pt, letterpaper]{ltxguide}
\usepackage[margin=2.5cm]{geometry}
\usepackage[english]{babel}
\usepackage[latin1]{inputenc}
\usepackage{amssymb}
\usepackage{bm}
\usepackage{microtype}
\usepackage{rotating}
\usepackage{econometrics}
\usepackage[authoryear,round]{natbib}
\usepackage{booktabs}
\usepackage{url}
\usepackage{indentfirst}                 
\usepackage{amssymb,amsmath,amsfonts,eurosym,geometry,ulem,graphicx,caption,color,setspace,sectsty,comment,caption,natbib,pdflscape,subfigure,array}
\usepackage{xcolor}
\usepackage{colortbl,booktabs}
\usepackage{multirow}
\usepackage{lettrine}
\usepackage[bottom]{footmisc} 
\usepackage{enumitem}

\usepackage{amsthm}
\usepackage{geometry}
\usepackage{algorithm}  
\usepackage{algcompatible}
\numberwithin{equation}{section}
\usepackage{caption}
\usepackage{multirow}
\usepackage{caption,tabularx,booktabs}
\usepackage{threeparttable}
\usepackage[colorlinks,
urlcolor=black,
            linkcolor=red,       
            anchorcolor=blue,  
            citecolor=blue,        
            ]{hyperref}
\newtheorem{theorem}{Theorem}

\newtheorem{lemma}{Lemma}

\newtheorem{definition}{Definition}

\newtheorem{remark}{Remark}[section]

\newcolumntype{L}[1]{>{\raggedright\let\newline\\arraybackslash\hspace{0pt}}m{#1}}
\newcolumntype{C}[1]{>{\centering\let\newline\\arraybackslash\hspace{0pt}}m{#1}}
\newcolumntype{R}[1]{>{\raggedleft\let\newline\\arraybackslash\hspace{0pt}}m{#1}}

\usepackage{xcolor}

\begin{document}
\newgeometry{left=2.1cm,right=2.1cm,top=1.5cm, bottom=2cm}
\begin{titlepage}
\title{Estimation and Testing of Forecast Rationality\\ with Many Moments}
\author{ Tae-Hwy Lee\vspace{0.5cm}\thanks{Department of Economics, University of California, Riverside, CA 92521, USA. E-mail: {taelee@ucr.edu}.} \ \ \ \ \  \ \quad Tao Wang\thanks{Department of Economics and Department of Mathematics and Statistics (by courtesy), University of Victoria, Victoria,\\ BC V8W 2Y2, Canada. E-mail: {taow@uvic.ca}.\\  We are deeply grateful to Editor William A. Barnett, an anonymous Associate Editor, and two anonymous reviewers for their constructive comments, leading to the substantial improvement of the paper. We would also like to thank Aman Ullah, Dingli Wang, and seminar participants at the University of Victoria, Purdue University, and the 2024 Annual Meetings of the Canadian Economics Association for their helpful comments. Tao Wang's research is supported by the Social Sciences and Humanities Research Council of Canada Insight Development Grant
(430-2023-00149) and the Natural Sciences and Engineering Resea-\\rch Council of Canada Discovery Grant (RGPIN-2025-04185 and DGECR-2025-00343).\looseness=-1}}

\maketitle
\begin{abstract}
We in this paper employ a penalized moment selection procedure to identify valid and relevant moments for estimating and testing forecast rationality within the flexible loss framework proposed by \cite{EKT05}. We motivate the selection of moments in a high-dimensional setting, outlining the fundamental mechanism of the penalized moment selection procedure and demonstrating its implementation in the context of forecast rationality, particularly in the presence of potentially invalid moment conditions. The selection consistency and asymptotic normality are established under conditions specifically tailored to economic forecasting. Through a series of Monte Carlo simulations, we evaluate the finite sample performance of penalized moment estimation in utilizing available instrument information effectively within both estimation and testing procedures. Additionally, we present an empirical analysis using data from the Survey of Professional Forecasters issued by the Federal Reserve Bank of Philadelphia to illustrate the practical utility of the suggested methodology. The results indicate that the proposed post-selection estimator for forecaster's attitude performs comparably to the oracle estimator by efficiently incorporating available information. The power of rationality and symmetry tests leveraging penalized moment estimation is substantially enhanced by minimizing the impact of uninformative instruments. For practitioners assessing the rationality of externally generated forecasts, such as those in the Greenbook, the proposed penalized moment selection procedure could offer a robust approach to achieve more effici-\\ent estimation outcomes.\looseness=-1 
\vspace{0.1in}

\vspace{0.08in}

\bigskip
\end{abstract}
\setcounter{page}{0}
\thispagestyle{empty}
\end{titlepage}
\pagebreak \newpage

\setlength\parindent{2em}
\allowdisplaybreaks[4]

\restoregeometry
\section{Introduction} \label{sec:introduction}
\noindent Forecasting is an essential technique in economics, statistics, and other sciences, serving as a foundational tool for decision-making under uncertainty. For forecast producers, optimal forecasting involves using all available information to minimize expected loss given a specified loss function. For forecast users, however, the focus shifts to evaluating the rationality of forecasts produced by others, such as government agencies (e.g., the Greenbook), even when the forecast producer's loss function is unknown. \cite{EKT05} (EKT (2005)) address this by proposing a framework that tests forecast rationality under a broad class of loss functions, which encompasses commonly used loss functions as special cases (Figure \ref{fig1}). In this framework, the family of loss functions is indexed by a single unknown parameter $\alpha$, which holds economic significance by reflecting the forecaster's objective or attitude toward forecast errors. A number of studies have further developed and applied the EKT (2005) framework for forecast evaluation under asymmetric loss. \cite{PT12} extend the flexible loss model to capture forecaster heterogeneity and information rigidity, while \cite{EK16} focus on forecast optimality under general loss functions in time series environments. \cite{I16} and \cite{SKG21} investigate instrument validity and model misspecification in rationality testing, highlighting the sensitivity of test results to instrument choice. \cite{BP22} propose robust approaches for forecast testing under asymmetry, and \cite{DPS24} introduce tools to infer forecaster preferences from observed forecasts. While these studies emphasize modeling and testing under flexible loss, our work will complement and extend this literature by developing a penalized approach that performs automatic moment selection, addressing the dual challenges of instrument validity and relevance, particularly important in high-dimensional forecasting environments.\looseness=-1

\vspace{0.2cm}

\begin{figure}[h]
{\centering
\includegraphics[scale=0.195]{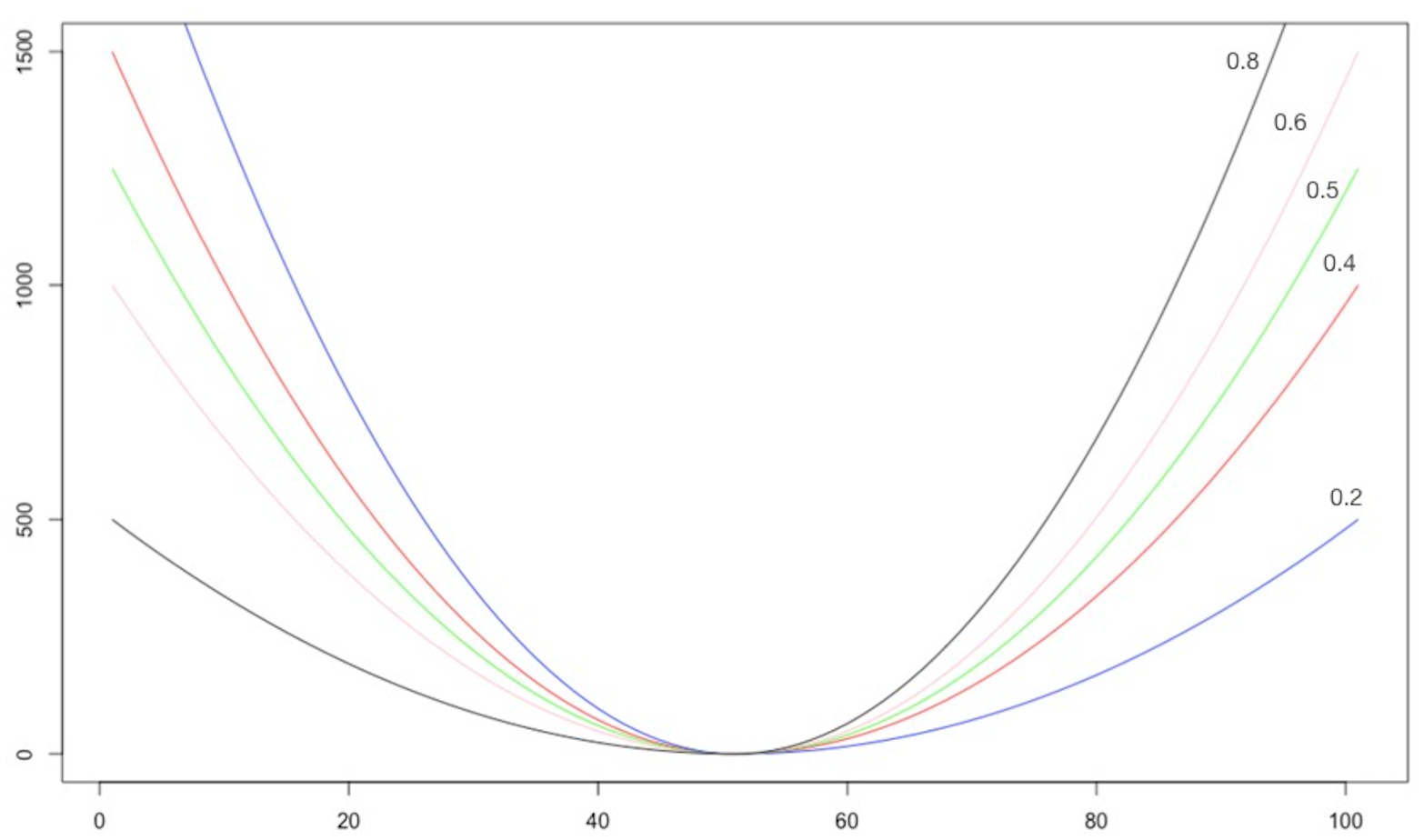} 
\quad \quad \includegraphics[scale=0.2]{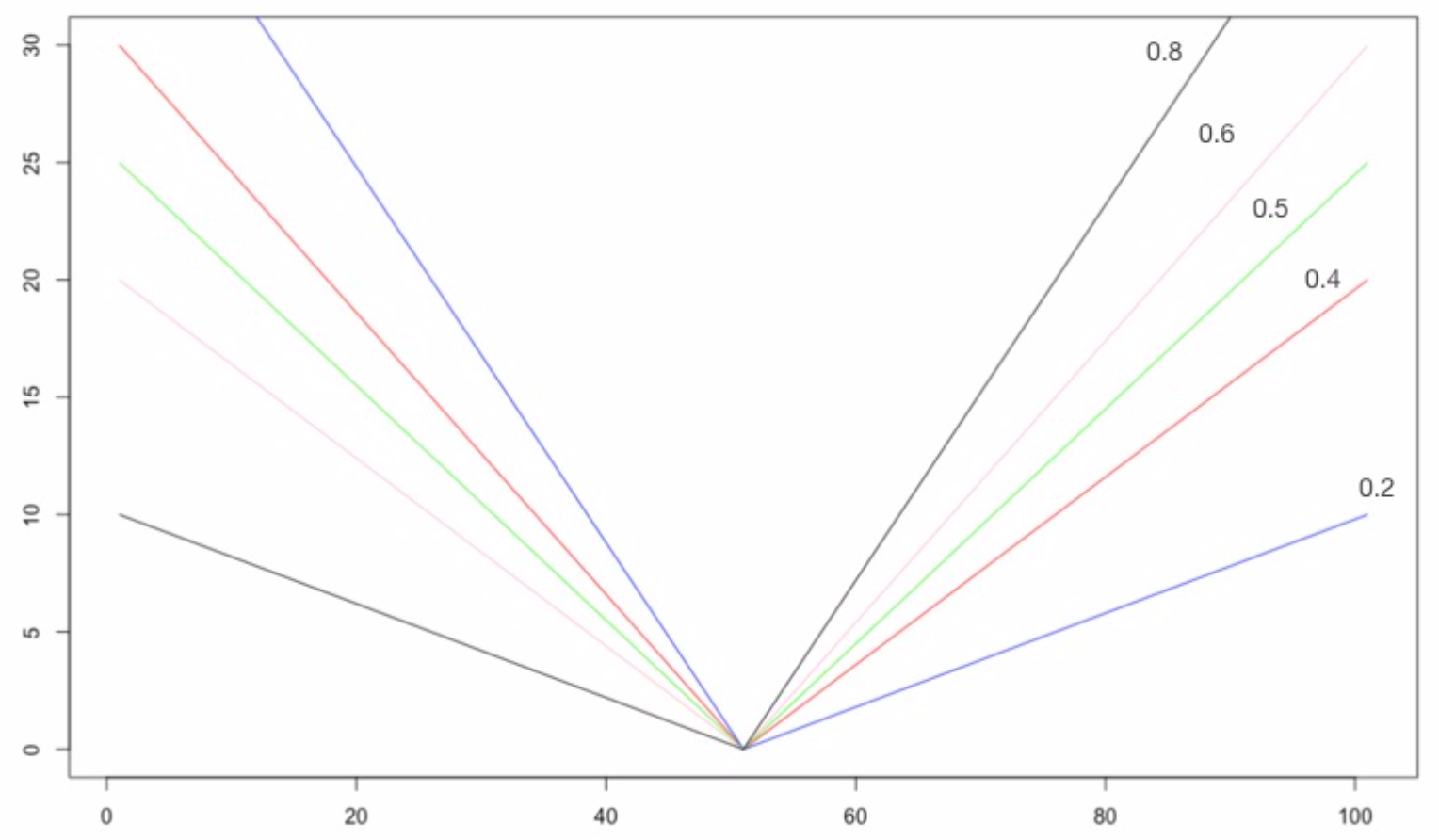} 
\caption{Loss Function with Different Values of $\alpha$ (Left: $p=2$; Right: $p=1$)}}
\vspace{0.2cm}
{\noindent {{\bf{\textit{Note}}}: The plot illustrates how changes in the asymmetry parameter $\alpha$ affect univariate loss functions. Even small deviations from the symmetric loss value ($\alpha$=0.5) lead to substantial loss differences. For instance, with $\alpha=0.4$, the loss ratio between positive and negative forecast errors is $\alpha/(1-\alpha)=2/3$, resulting in a loss differential of approximately 33\%.\looseness=-1  }}
\end{figure}\label{fig1}

At the core of the EKT (2005) framework lies a definition of forecast rationality based on optimal forecast generation, wherein rational forecasts are those that minimize expected losses under the forecaster's subjective, and possibly asymmetric, loss function. EKT (2005) establish conditions under which the forecaster's attitude parameter $\alpha$ can be identified and propose a $J$-test for overidentification to assess forecast rationality within a generalized method of moments (GMM) setting that accommod-\\ates asymmetric loss. The loss function introduced by EKT (2005) is specified as follows\looseness=-1  
\vspace{-0.1cm}
\begin{equation}\label{eq1}
L\left(p,\alpha, e_{t+h}\right) \equiv\left[\alpha+(1-2 \alpha) \mathbb{I}\left(e_{t+h}<0\right)\right]\lvert e_{t+h}\rvert^{p},
\vspace{-0.1cm}
\end{equation}
in which $e_{t+h}$ stands for the forecast error, $\mathbb{I}(\cdot)$ is an indicator function that equals one when $e_{t+h}<0$ (overestimation) and zero otherwise, \(p\) is a positive exponent determining the curvature of the loss function, \(\alpha \in (0,1)$ is an asymmetry parameter, defining the degree of the forecaster's tolerance for errors\\ in different directions, and $h \geq 0$ is an integer variable that measures the forecast horizon.\looseness=-1

In accordance with the optimization problem \eqref{eq1}, the appropriate action that forecast producer takes should satisfy the forecast optimality condition\looseness=-1
\vspace{-0.1cm}
\begin{equation}
\mathbb{E}\left(W_{t} \left[\mathbb{I}\left(e_{t+h}<0\right)-\alpha_{0}\right] \lvert e_{t+h}\rvert^{p_{0}-1}\right)=0,
\vspace{-0.1cm}
\end{equation}
where $\alpha_0$ and $p_0$ represent the true values of $\alpha$ and $p$, respectively, \(W_{t}\) denotes the information (instrument) set available to the forecaster at time \(t\), and the optimal forecast error forms a martingale difference sequence relative to $W_t$; see \cite{EK08, EK16}. However, forecast users may not have access to the full information set \(W_{t}\). To estimate \(\alpha,\) based on the orthogonality of martingale\\ differences, it suffices to have a subset of instruments \(V_{t} \subset W_{t}\) such that\looseness=-1 
\vspace{-0.1cm}
\begin{equation}\label{eq3}
\mathbb{E}\left(V_{t} \left[\mathbb{I}\left(e_{t+h}<0\right)-\alpha_{0}\right] \lvert e_{t+h}\rvert^{p_{0}-1}\right)=0,
\vspace{-0.1cm}
\end{equation}
which is sufficient to identify $\alpha$ even under model misspecification.

The choice of instrument variables $V_t$ is critical for accurately estimating and testing forecast rationality. Using the EKT (2005) approach to uncover the forecaster's attitude parameter $\alpha$ generally involves a broad set of instruments to capture all available information that the forecast producer might use. In practice, however, the accessible information set \(V_{t}\) can be extensive, often including potential nonlinear transformations of instruments. For example, with six instruments, there are $2^6-1=63$ possible subsets of instruments, and incorporating nonlinear transformations (e.g., polynomial terms) greatly increases this number. Such expansion can result in a singular weighting matrix in the GMM estimation of $\alpha$. While economic rationale could sometimes guide instrument selection, in many instances, instrument choice may be arbitrary \citep{KM08, I16}. This can lead to biased estimators, especially if the chosen instruments do not match those actually used by the forecast producer, with bias worsening as the number of moments $d=\dim(V_t)$ increases. Additionally, estimation outcomes are highly sensitive to instrument choice: invalid instruments lead to inconsistent estimates of $\alpha$, while redundant instruments add finite sample bias without enhancing efficiency. Therefore, to ensure consistent and efficient estimation of $\alpha$ and improve the power of rationality and symmetry tests, identifying valid and relevant instruments (moment conditions) is essential. To the best of our knowledge, the EKT (2005) framework has not yet been integrated with moment selection methods to address the uncertain validity and relevance of moments when many potential instruments are available. This paper aims to fill this gap in the literature by identifying instruments that are both valid and\\ relevant to estimate $\alpha$, thereby shedding light on the forecaster's preferences and loss attitudes.\looseness=-1

Considerable research has been dedicated to selecting moment conditions in \eqref{eq3} when faced with numerous potential instruments. For instance, \cite{A99} and \cite{AL01} propose a moment selection criterion that combines the $J$-test statistic for overidentifying restrictions with a penalty for the number of moment conditions. \cite{HPS03} develop selection criteria for unconditional moment models using empirical likelihood statistics, while \cite{L13} suggests a GMM shrinkage method by adding a penalty function to the GMM criterion for moment selection. Further, \cite{CHL18} introduce the adaptive elastic net GMM estimator tailored for high-dimensional models with potentially invalid moments. However, these methods primarily focus on moment validity, assuming all moments are relevant. To address relevance, \cite{A02} and \cite{I06} employ a bootstrap approach grounded in Edgeworth expansions, while \cite{HIJS07} present an entropy-based moment selection criterion. \cite{NB09} propose an $L_2$-Boosting technique for selecting relevant instruments, and \cite{LC16} introduces a LASSO-based procedure using the $L_1$ penalty to identify informative moments. Nonetheless, each of these approaches operates under the assumption that all candidate moments are valid, an assumption that is problematic for any robust forecast evaluation method. Building on the shrinkage procedure proposed by \cite{L13}, \cite{CL15} explore the simultaneous selection of valid and relevant moments using a penalized GMM estimation approach, referred to as P-GMM. This method introduces an innovative penalty that accounts for both moment validity and relevance, allowing for adaptive estimation. Expanding on this, \cite{BCCHK18} recommend a regularized GMM framework, constructing moment equations \(M(\theta ; \eta)=0\) for the target parameter \(\theta \), conditional on the nuisance parameter \(\eta \), such that the true parameter values satisfy \(M\left(\theta_{0} ; \eta_{0}\right)=0\). These two studies demonstrate that P-GMM can effectively select both valid and rele-\\vant moment conditions simultaneously, enhancing the accuracy of GMM estimation.\looseness=-1

In the EKT (2005) framework, the primary objective is not to estimate regression coefficients, and information on the specific variables used by forecast producers is often limited. Consequently, traditional approaches that prioritize estimating target coefficients are not directly applicable for selecting the valid and relevant moments necessary to identify the forecaster's attitude parameter $\alpha$. The P-GMM approach introduced by \cite{L13} and \cite{CL15}, which incorporates a penalty within the standard GMM criterion, offers a promising solution for selecting moment conditions. In this paper, we adapt the P-GMM technique specifically within the EKT (2005) framework, enabling the selection of moment conditions for estimating and testing forecast rationality under flexible, asymmetric loss functions. This adaptation is not a simple application of \cite{CL15}'s method but an extension tailored to the distinct requirements of the EKT (2005) context. We rigorously demonstrate that the P-GMM approach achieves moment selection consistency in this new setting, effectively identifying all valid and relevant moment conditions without requiring detailed knowledge of the forecast producer's information set. This leads to a more efficient post-selection estimation process, where the P-GMM estimator asymptotically attains the oracle property under consistent moment selection, matching the efficiency of an oracle GMM estimator based on all valid and relevant moment conditions. By extending P-GMM to address the unique challenges of the EKT (2005) framework, this paper provides a theoretically sound and practically relevant approach for efficient moment selection\\ that accommodates the inherent asymmetries in the information available to forecast users.\looseness=-1

To elucidate the statistical properties of P-GMM estimation in assessing forecast rationality, we conduct a series of Monte Carlo simulations encompassing both linear and nonlinear dependencies between the forecast user's information set and forecast error. The simulation results indicate that the P-GMM procedure provides an unbiased and more efficient estimator of the forecaster's attitude parameter $\alpha$, compared to cases using all (valid) moment conditions. Notably, P-GMM estimation delivers post-selection outcomes that approximate oracle estimators by efficiently incorporating available information. This approach also reduces standard errors and enhances the power of the $J$-test statistics by mitigating the influence of irrelevant instruments. These findings underscore P-GMM's practical utility in scenarios where forecast users lack prior knowledge of instrument validity, allowing for effective moment selection within the EKT (2005) framework to achieve consistent and efficient estimation. We then apply the suggested P-GMM estimation to data from the Survey of Professional Forecasters (SPF) issued by the Federal Reserve Bank of Philadelphia to further elucidate its practical value. The empirical results emphasize the importance of moment selection, as estimates of the forecaster's attitude parameter $\alpha$ reveal distinct asymmetries across different variables. For real GNP/GDP and the price index, $\alpha$ estimates are close to 0.5, reflecting symmetric error penalties under quadratic loss, while estimates for consumption and investment deviate significantly from 0.5, suggesting an asymmetric loss function with higher weights on negative forecast errors. Additionally, P-GMM estimation yields more efficient estimates, as indicated by smaller standard errors and enhanced estimation precision. The results suggest that selecting valid and relevant moments can eliminate inefficiencies associated with\\ using all instruments and enhance the accuracy of the forecast rationality test.\looseness=-1

\indent The structure of this paper is as follows. Section \ref{sec:EKT} addresses the moment selection challenges within the EKT (2005) framework, motivating the development of the proposed estimation procedure. Section \ref{sec:selection} introduces the P-GMM estimation method, which selects valid and relevant moment conditions in the EKT (2005) context, along with demonstrating moment selection consistency and asymptotic normality. Section \ref{sec:numerical} provides simulation examples to illustrate the statistical properties of the proposed P-GMM estimation and testing approach for forecast rationality. Section \ref{sec:empirl} presents a real data analysis using the SPF dataset to showcase the practical performance of the P-GMM estimation.\\ We conclude this paper in Section \ref{sec:conclusion}. All technical proofs are listed in the Appendix.\looseness=-1

\section{Moment Selection in Forecast Rationality}\label{sec:EKT}
\noindent This section explores the critical role of moment selection in testing forecast rationality within the EKT (2005) framework, highlighting both its importance and challenges, and motivating the proposed estimation procedure. Forecast rationality in the EKT (2005) framework is assessed through an asymmetric loss function, capturing a forecaster's preferences via an attitude parameter $\alpha$ that reflects asymmetric penalties for over- and under-predictions. Effective estimation of $\alpha$ requires that we identify the optimal forecast model, along with selecting moments that capture all relevant and valid information.\looseness=-1

Consider a stochastic process $ X =\{X_{t} : \Omega \rightarrow \mathbb{R}^{m+1}, m \in \mathbb{N}, t=1, \dots, T\}$ on a probability space \((\Omega, \mathcal{F}, P)\), where \(\mathcal{F}=\left\{\mathcal{F}_{t}, t=1,\dots,T \right\} \) and \(\mathcal{F}_{t}\) is the \(\sigma\)-field generated by \(\mathcal{F}_{t}=\sigma\left\{X_{s}, s \leqslant t\right\}\). Let \(Y_{t}\) be the continuous variable of interest in the observed vector $X_t$. The $h$-step-ahead forecast $f_{t+h}$ of $Y_{t+h}$ depends on an information set \(\mathcal{F}_{t}\) and is modeled as \(f_{t+h} = \theta^{\prime} W_{t}\), where \(\theta \)  is an unknown vector of parameters and \(W_{t}\) contains \(\mathcal{F}_{t}\)-measurable variables. EKT (2005) propose a generalized loss function,\\ $L(p, \alpha, \theta)$, that accommodates asymmetry in the penalization of forecast errors\looseness=-1
\vspace{-0.1cm}
\begin{equation}\label{eq2.1}
L(p, \alpha, \theta) \equiv\left[\alpha+(1-2 \alpha) \mathbb{I}\left(Y_{t+h}-f_{t+h}<0\right)\right]\lvert Y_{t+h}-f_{t+h}\rvert^{p},
\vspace{-0.1cm}
\end{equation}
where \(Y_{t+h}-f_{t+h}=Y_{t+h}-\theta^{\prime} W_{t}\) represents the forecast error \(e_{t+h}\). For analytical simplicity, EKT (2005) assume that the true value of the loss shape parameter $p_0$ is known and deliberate on the estimation of forecaster's attitude parameter $\alpha$. Given \(p_{0}\) and the true value \(\alpha_{0}\), forecast rationality implies that the forecast $f_{t+h}^{*}=\theta^{* \prime} W_{t}$ is optimal when it minimizes expected loss. Formally, this optimality\\ condition is represented by $\theta^{*}=\arg \min _{\theta \in \Theta}$ $\mathbb{E}\left\{L\left(p_{0}, \alpha_{0}, \theta\right)\right\} $, leading to the moment condition\looseness=-1
\vspace{-0.05cm}
\begin{equation}\label{eq5}
\mathbb{E} \left(W_{t} \left[\mathbb{I}\left(Y_{t+h}-f_{t+h}^{*}<0\right)-\alpha_{0}\right]\lvert Y_{t+h}-f_{t+h}^{*}\rvert^{p_{0}-1}\right)=0.
\vspace{-0.05cm}
\end{equation}
Note that \eqref{eq5} can be considered as a moment condition \( \mathbb{E}[W_{t}(\mathbb{I}(e_{t+h}^{*}<0)-\alpha_{0})\lvert e_{t+h}^{*}\rvert^{p_{0}-1}]=0\), where \(e_{t+h}^{*} = Y_{t+h}-f_{t+h}^{*}=Y_{t+h}-\theta^{* \prime} W_{t}\) is the forecast error given $f^*_{t+h}$. It suggests that if the forecasts are optimal, then any information $W_t$ must be correctly included in $f_{t+h}^{*}$, so that $W_{t}(\mathbb{I}(e_{t+h}^{*}<0)-\alpha_{0})$ is orthogonal to $\lvert Y_{t+h}-\theta^{* \prime} W_{t} \rvert^{p_0-1}$. For accurate estimation of \(\alpha\), this condition highlights the importance\\ of selecting the right moments that capture all relevant information in $W_t$.\looseness=-1

In practice, not all variables in $W_t$ may be observable or known to the forecast user. Forecast users typically rely on an observed $d$-vector $V_t \subset W_t$ to approximate the information set, leading to a modif-\\ied moment condition\looseness=-1
\vspace{-0.05cm}
\begin{equation}\label{eq6}
\mathbb{E} (V_{t} [\mathbb{I}(Y_{t+h}-f_{t+h}^{*}<0)-\alpha_{0}]\lvert Y_{t+h}-f_{t+h}^{*}\rvert^{p_{0}-1})=\mathbb{E}(C_t-\alpha_0 B_t)=0,
\vspace{-0.05cm}
\end{equation}
where \( C_t=V_{t} \mathbb{I}(Y_{t+h}-f_{t+h}^{*}<0) \lvert Y_{t+h}-f_{t+h}^{*}\rvert^{p_{0}-1}\) and \(B_t=V_{t} \lvert Y_{t+h}-f_{t+h}^{*}\rvert^{p_{0}-1}\). Thenceforth, $\alpha_0$ can be solved by minimizing the quadratic norm of \eqref{eq6} such that $\min _{\alpha} Q(\alpha)=\mathbb{E}(C_t-\alpha B_t)' S^{-1} \mathbb{E}(C_t-\alpha B_t)$, in which $S=\mathbb{E}((C_t-\alpha B_t)(C_t-\alpha B_t)')=\mathbb{E}[V_{t} V_{t}^{\prime}\left(\mathbb{I}\left(Y_{t+h}-f_{t+h}^{*}<0\right)-\alpha_{0}\right)^{2}\lvert Y_{t+h}-f_{t+h}^{*}\rvert^{2 p_{0}-2}]$ is a positive definite weighting matrix. Solving for $\alpha_0$ under this condition is non-trivial because the inclusion of irrelevant or invalid instruments in $V_t$ can distort the estimation, introducing inefficiencies and bias. Therefore, moment selection is crucial to isolating only those instruments that are both valid\\ and relevant for consistent estimation of $\alpha$.\looseness=-1

Given a sample of instrument vectors \(\left(V_{1}^{\prime}, \ldots, V_{T}^{\prime}\right)^{\prime}\) with \(T\) observations, an estimator for $\alpha_0$ is\\ obtained as\looseness=-1
\vspace{0.055cm}
\begin{equation}\label{eq2.6}
\hat{\alpha}_{T}=(\hat{B}_{T}^{\prime} \hat{S}_{T}^{-1} \hat{B}_{T})^{-1}(\hat{B}_{T}^{\prime} \hat{S}_{T}^{-1} \hat{C}_{T}),
\vspace{0.055cm}
\end{equation}
where $\hat{B}_{T}=1/T \sum_{t=1}^{T} V_{t}\lvert Y_{t+h}-{f}^*_{t+h}\rvert^{p_{0}-1}$, $\hat{C}_{T}={1}/{T} \sum_{t=1}^{T} V_{t} \mathbb{I}(Y_{t+h}-{f}^*_{t+h}<0)\lvert Y_{t+h}-{f}^*_{t+h}\rvert^{p_{0}-1}$, and $\hat{S}_{T} \equiv {1}/{T} \sum_{t=1}^{T} [V_{t} V_{t}^{\prime}\left(\mathbb{I}\left(Y_{t+h}-f_{t+h}^{*}<0\right)-\hat{\alpha}_{T}\right)^{2}\lvert Y_{t+h}-f_{t+h}^{*}\rvert^{2 p_{0}-2}]$ is a consistent estimate of \(S\). The accuracy of $\hat{\alpha}_{T}$ depends heavily on the selection of valid instruments in $V_t$, as each set of instruments produces different estimates and asymptotic variances. In practice, however, forecast users may not know which variables in $V_t$ are valid, underscoring the need for an effective estimation procedure to\\ assess instrument validity and avoid bias.\looseness=-1

Following the estimator construction, we have asymptotic normality (EKT (2005) $Proposition$ 4) for $\hat{\alpha}_{T}$, i.e., $T^{\frac{1}{2}}\left(\hat{\alpha}_{T}-\alpha_{0}\right) \stackrel{d}{\rightarrow} \mathcal{N}(0,(B^{\prime} S^{-1} B)^{-1})$ with $B=\mathbb{E}(B_t)$, which can be utilized to test whether $\hat{\alpha}_{T}$ differs significantly from $\alpha_{0}$. For rationality testing (as outlined in EKT (2005) $Corollary$ 5), we can conduct a joint test of forecast rationality and flexible loss specification when $d >1$ instru-\\ments are available, using the overidentification test statistic\looseness=-1
\vspace{-0.14cm}
\begin{equation}\label{eq11}
J_{T}\left(\hat{\alpha}_{T}\right)=T \times \hat{Q}_{T}\left(\hat{\alpha}_{T}\right)=T \times \hat{A}_{T}\left(\hat{\alpha}_{T}\right)^{\prime} \hat{S}_{T}^{-1}\left(\hat{\alpha}_{T}\right) \hat{A}_{T}\left(\hat{\alpha}_{T}\right) \stackrel{d}{\rightarrow} \chi_{d-1}^{2},
\vspace{-0.14cm}
\end{equation}
where \(\hat{A}_{T}\left(\hat{\alpha}_{T}\right)=\hat{C}_{T}-\hat{\alpha}_{T} \hat{B}_{T}\) and a large $J_{T}(\hat{\alpha}_T)$ suggests rejection of forecast rationality. In the presence of a single instrument, overidentification testing is not feasible, yielding a unique closed-form solu-\\tion for $\hat{\alpha}_T$. For testing loss symmetry, we can fix $\hat{\alpha}_T=0.5$, resulting in\looseness=-1
\vspace{-0.14cm}
\begin{equation}\label{eq12}
J_{T}\left(0.5 \right)=T \times \hat{Q}_{T}\left(0.5\right)=T \times \hat{A}_{T}\left(0.5\right)^{\prime} \hat{S}_{T}^{-1}\left(\hat{\alpha}_{T}\right) \hat{A}_{T}\left(0.5\right)\stackrel{d}{\rightarrow} \chi_{d}^{2}
\vspace{-0.14cm}
\end{equation}
with $d$ degrees of freedom as no parameter estimation is involved. As suggested in EKT (2005), the use of $\hat{S}_{T}^{-1}\left(\hat{\alpha}_{T}\right)$ can improve the finite sample power. The rejection of this test indicates asymmetry in loss preferences, provided the rationality hypothesis \eqref{eq11} is not rejected. While the conditional distribution of \eqref{eq12} is difficult to derive, following \cite{WL14}, we interpret $J_T(0.5)$ as a symmetry test following the selection of all valid and relevant moment conditions. These considerations underscore that the statistical properties of the GMM estimator $\hat{\alpha}_T$ depend heavily on the reliability of the moment conditions. Thus, incorporating information on both moment validity and relevance is advan-\\tageous for conducting robust rationality and symmetry tests.\looseness=-1

\begin{remark}\label{re2.1}
EKT (2005) assume that the forecasts are generated using a linear model $f_{t+h}=\theta' W_t$. In the case of a nonlinear model where $f_{t+h}=f(\theta, W_t)$ and $f(\cdot)$ is continuously differentiable, there ex-\\ists a parameter $\theta^*$ that satisfies\looseness=-1
\vspace{-0.16cm}
\begin{equation*}
\mathbb{E} \big(f^{*'}_{t+h} \big[\mathbb{I}\big(Y_{t+h}-f_{t+h}^{*}<0\big)-\alpha_{0}\big]\lvert Y_{t+h}-f_{t+h}^{*}\rvert^{p_{0}-1}\big)=0,
\vspace{-0.16cm}
\end{equation*}
where $f^{*'}_{t+h}=\partial f(\theta^*, W_t)/ \partial \theta^*$. This allows for the use of $V_{v,t}$, a subvector of $f^{*'}(\theta^*, W_t)$, to conduct the rationality test. In both linear and nonlinear cases, the selection of valid and relevant moments is essential for accurate estimation and testing, as the choice of moments directly influences the efficiency and consistency of the parameter estimates; see simulations in Section \ref{sec:numerical}. However, the nonlinear case poses an additional challenge, as $f^*(\theta^*, W_t)$ depends on both $\theta^*$ and $W_t$, while the true $\theta^*$ used by the\\ forecast user is typically unknown. Consequently, this paper focuses exclusively on the linear case.\looseness=-1
\end{remark}

\vspace{-0.05cm}

\indent Rationality tests under asymmetric loss functions have received extensive attention in forecasting literature, as traditional reliance on the squared error loss function may not fully capture the diverse preferences of forecasters; see \cite{CD97}, \cite{C98}, \cite{PT07}, \cite{EKT05, EKT08}, \cite{P20}, \cite{BS24}, among others. The EKT (2005) framework is particularly notable for its flexibility in accommodating various asymmetric loss functions, enabling more tailored tests of forecast optimality across diverse forecasting contexts. However, a critical underlying assumption in EKT (2005) is that the forecast user has access to a comprehensive set of information, as captured by the moment condition in \eqref{eq5}. In practice, forecast users often have incomplete information compared to forecasters, making \eqref{eq6} particularly useful,\\ as it reflects a more realistic scenario where users cannot access all forecaster information.\looseness=-1

\begin{figure}[h]
{\centering
\includegraphics[scale=0.31]{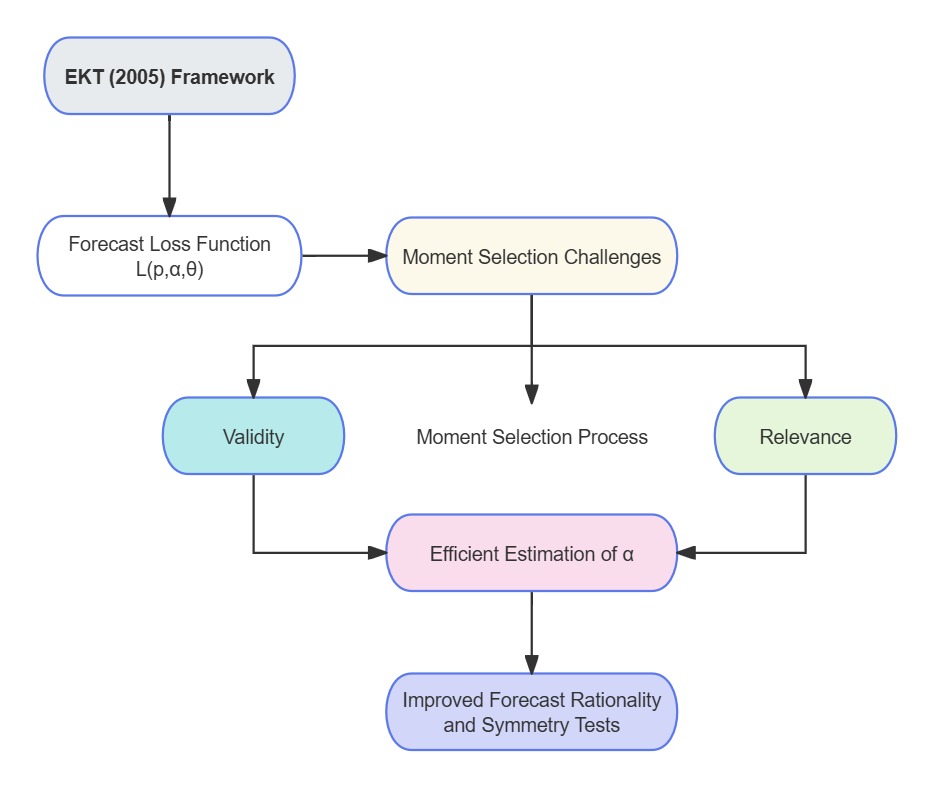}
\vspace{-0.3cm}
\caption{Moment Selection in EKT (2005) Framework}\label{figggg} }
\vspace{0.25cm}
\noindent {{\bf{\textit{Note}}}: The plot highlights the importance of selecting valid and relevant moments in the EKT (2005) framework to ensure efficient estimation of the forecaster's attitude parameter $\alpha$, leading to more reliable and accurate forecast rationality tests. By carefully choosing moments, we can enhance the robustness of the model and improve the accuracy of subsequent analyses.\looseness=-1} 
\vspace{-0.15cm}
\end{figure}

Despite having access to an observed set of instruments $V_t$, it remains challenging to determine how forecasting institutions incorporate these variables into their models and which instruments will yield the most efficient estimate of the forecaster's attitude parameter $\alpha$. Data-driven selection methods, such as those proposed by \cite{DN01}, which focus on isolating valid instruments within a predefined subset, provide partial solutions but remain limited. When many potential instruments are available, and their validity is uncertain, selecting irrelevant or invalid instruments can lead to biased estimates and unreliable rationality testing outcomes. This highlights the importance of rigorously assessing instrument validity and relevance, which is the primary focus of this paper (see Figure \ref{figggg}). To address these challenges, we in the following section employ a P-GMM estimation approach within the EKT (2005) framework, enabling systematic selection of all valid and relevant instruments,\\ thus ensuring consistent and efficient estimation and testing under flexible loss preferences.\looseness=-1

\section{Penalized Moment Selection: Validity and Relevance} \label{sec:selection}
\begin{figure}[h]
{\centering
\includegraphics[scale=0.38]{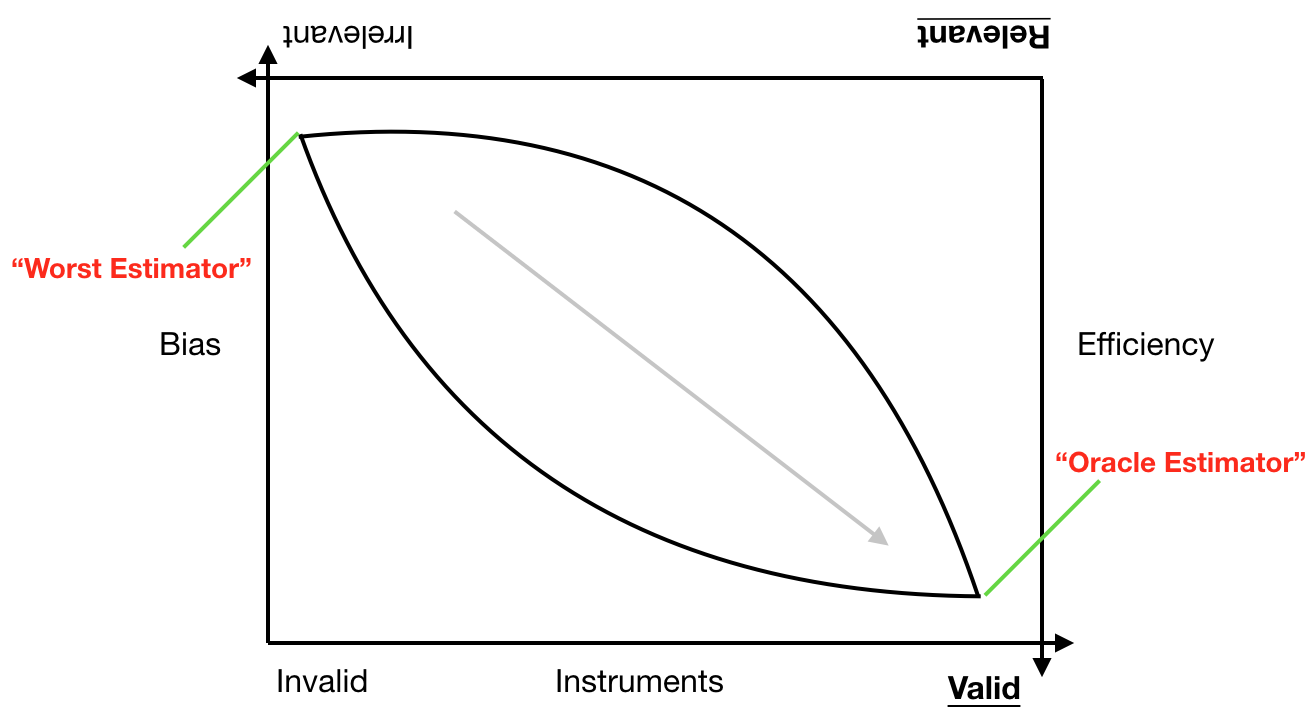}
\caption{Roadmap of Instrument Selection}\label{figg2} } 
\vspace{0.3cm}
\noindent {{\bf{\textit{Note}}}: The instruments could be divided into four types: valid and invalid, relevant and irrelevant. To reduce bias, valid instruments must be selected. For obtaining efficiency, relevant instruments should be chosen. To get the unbiased and efficient estimation of the forecaster's attitude parameter $\alpha$, we need to choose valid and relevant instruments, which is close to  ``Oracle Estimator."\looseness=-1 }
\end{figure}

\noindent The estimation and testing of forecast rationality under flexible loss functions are of significant interest for forecast users, as these processes are critical both theoretically and empirically. A key concern arises when forecasts fail to efficiently incorporate the information in $V_t$; in such cases, the estimator of $\alpha$ would differ substantially across moment conditions, leading to a rejection in the overidentification test. Despite the practical importance of accurate moment selection, empirical applications currently lack standardized guidelines for choosing appropriate instruments. Determining the suitable (valid and relevant) instruments to reveal the forecaster's attitude parameter $\alpha$ and to improve estimation efficiency remains an open question in the context of EKT (2005). To illustrate the motivation behind the\\ methodology proposed in this paper, consider a set of potentially misspecified moment conditions\looseness=-1
\vspace{-0.1cm}
\begin{equation}
\mathbb{E} \left(V_{t} \left[\mathbb{I}\left(Y_{t+h}-f_{t+h}^{*}<0\right)-\alpha_{0}\right]\lvert Y_{t+h}-f_{t+h}^{*}\rvert^{p_{0}-1}\right) \overset{?}{=} 0,
\vspace{-0.1cm}
\end{equation}
where the instruments $V_t$ satisfy equality only for certain moments, indicated by the notation ``$\overset{?}{=}$''. Including these misspecified moments in estimation risks producing inconsistent results for $\alpha$. To address this, we employ the P-GMM estimation method \citep{L13, CL15}, which can select valid and relevant moment conditions simultaneously to obtain not only an unbiased but also a more\\ efficient estimator of $\alpha$ (Figure \ref{figg2}).\looseness=-1

\vspace{-0.1cm}

Practically, \eqref{eq6} serves as the fundamental criterion for selecting valid moment conditions. Specifically, if an instrument fails to satisfy \eqref{eq6}, it cannot be used for estimation purposes. Once the set of valid instruments is determined, the next step is to identify the relevant instruments by examining the variance of the resulting estimators. A set of moment conditions that produces a smaller variance, indicating that it incorporates more valuable information, will be considered relevant, with $(B'S^{-1}B)^{-1}$ serving as the primary metric for this relevance. These two selection steps, focused on validity and relevance, are critical and widely emphasized in the literature on instrument or moment selection; see \cite{LSL10}. P-GMM estimation streamlines these two steps by incorporating both validity and relevance assessments through a penalized approach, attaching an appropriate penalty to the standard GMM criterion. This method offers a systematic approach to refine moment selection, balancing bias and efficiency. Before presenting the specifics of P-GMM estimation within the EKT (2005) frame-\\work, we first define the concepts of valid and relevant instruments as applied in this paper.\looseness=-1

\begin{definition}\label{def1}
(Validity) An instrument $V_{i,t}$ is claimed as a valid instrument in terms of the estimation\\ of $\alpha$ if and only if it satisfies \eqref{eq6}, i.e., $
\mathbb{E}\left[V_{i,t}\left(\mathbb{I}\left(e_{t+h}^{*}<0\right)-\alpha_{0} \right)\lvert e_{t+h}^{*}\rvert^{p_{0}-1} \right]=0.$\looseness=-1
\end{definition}

\begin{definition}\label{def2}
(Relevancy) A valid instrument $V_{j,t}$ is said to be a relevant instrument in terms of the estimation of $\alpha$ given moment condition $
\mathbb{E}\left[(V_{j,t},V_{i,t})'\left(\mathbb{I}\left(e_{t+h}^{*}<0\right)-\alpha_{0} \right)\lvert e_{t+h}^{*}\rvert^{p_{0}-1} \right]=0$ if and only\\ if $
\hat{B}'_T(V_{j,t},V_{i,t})\hat{S}^{-1}_T(V_{j,t},V_{i,t}) \hat{B}_T(V_{j,t},V_{i,t})\geq \hat{B}'_T(V_{i,t})\hat{S}^{-1}_T(V_{i,t}) \hat{B}_T(V_{i,t}).$\looseness=-1
\end{definition}

To approach the moment selection process systematically, we categorize the instrument set $V_t$, with dimension $\dim(V_t)=d$, into two primary groups: ``Good'' and ``Doubt'' instruments. This categorization, represented as $V_t=(V_{G,t}, V_{D,t})'$, distinguishes the instruments $V_{G,t} \in \mathbb{R}^{k_0}$ (the ``Good'' instruments) from $V_{D,t} \in \mathbb{R}^{d-k_0}$ (the ``Doubt'' instruments) with $k_0<d$. For the ``Good'' group, we assume that the corresponding moment function, $g_G(V_{G,t}, \alpha)=C_{G,t}-\alpha B_{G,t} \in \mathbb{R}^{k_{0}}$, is directly usable for the identification of parameter $\alpha$ without further testing for validity or relevance, where \( C_{G,t}=V_{G,t}\mathbb{I}(Y_{t+h}-f_{t+h}^{*}<0) \lvert Y_{t+h}-f_{t+h}^{*}\rvert^{p_{0}-1}\) and \(B_{G,t}=V_{G,t} \lvert Y_{t+h}-f_{t+h}^{*}\rvert^{p_{0}-1}\). For instance, setting $k_0=2$ allows us to define $V_{G,t}=(1, Y_{t-1})'$, where the constant and lagged dependent variable are acknowledged as established valid and relevant instruments. To construct an optimal collection of moment conditions, we apply the framework  by \cite{L13} and \cite{CL15}, where the ``Doubt'' moment conditions are represented by $g_D(V_{D,t}, \alpha)=C_{D,t}-\alpha B_{D,t} \in \mathbb{R}^{d-k_{0}}$ and are subject to tests for both validity and relevance. Here, \( C_{D,t}=V_{D,t}\mathbb{I}(Y_{t+h}-f_{t+h}^{*}<0)  \lvert Y_{t+h}-f_{t+h}^{*}\rvert^{p_{0}-1}\) and \(B_{D,t}=V_{D,t} $ $\lvert Y_{t+h}-f_{t+h}^{*}\rvert^{p_{0}-1}\). The combined moment function, $g(V_t, \alpha)=C_t-\alpha B_t=[g_G(V_{G,t}, \alpha), g_D(V_{D,t}, \alpha)]' \in \mathbb{R}^{d}$, thus encompasses both the ``Good'' and ``Doubt'' sets of all available moments.\looseness=-1

To guide the selection of appropriate moments, we introduce a slackness parameter $\beta$ defined as $\beta=\mathbb{E}[g_D(V_{D,t}, \alpha)]$. This parameter allows us to refine the estimation within the P-GMM framework by adjusting for potential misspecifications in the ``Doubt'' instruments. Integrating this slackness par-\\ameter into \eqref{eq6}, we redefine the expectation condition as follows\looseness=-1
\begin{equation}\label{eq3.1}
\mathbb{E}\left(V_{t}\left[\mathbb{I}\left(Y_{t+h}-f_{t+h}^{*}<0\right)-\alpha_{0}\right]\lvert Y_{t+h}-f_{t+h}^{*}\rvert^{p_{0}-1}-\begin{bmatrix}
0\\
\beta_0
\end{bmatrix} \right)\equiv \mathbb{E} \begin{bmatrix}
g_G(V_{G,t}, \alpha_0)\\
g_D(V_{D,t}, \alpha_0)-\beta_0
\end{bmatrix} =0,
\end{equation}
where $\beta_0=\mathbb{E}[g_D(V_{D,t}, \alpha_0)]$. Subsequently, the instrument selection process is conducted within the set \({D}=\mathcal{A} \cup \mathcal{B}_{1} \cup \mathcal{B}_{0}\) with the cardinality $\lvert {D} \rvert=d-k_0$, in which $\mathcal{A}$ represents the set of valid and relevant moments, $\mathcal{B}_0$ indicates the set of valid but irrelevant ones, and $\mathcal{B}_1$ is the set of invalid moments. While both the moments in set $\mathcal{A}$ and $\mathcal{B}_0$ are valid, only those within $\mathcal{A}$ are considered relevant for efficient estimation. Leveraging this structure, the simultaneous estimation of $\alpha$ and $\beta$ is achieved through \eqref{eq3.1},\\ allowing us to derive robust estimates that incorporate the complete set of validated instruments.\looseness=-1

The P-GMM estimation approach achieves efficient estimation and selective moment selection by minimizing a penalized objective function designed to balance fit with the complexity of the model, thereby optimizing both parameter estimation and the relevance of the instruments employed. Specifi-\\cally, this is expressed as\looseness=-1
\vspace{-0.1cm}
\begin{equation}\label{eq14}
{\hat{\alpha}}_{0}=\underset{\alpha}{\arg \min }\bigg[{\overline{g}_T}(\alpha, \beta)^{\prime} S^{-1}_T {\overline{g}_T}(\alpha, \beta)+\lambda_T \sum_{\ell \in \mathcal{A} \cup \mathcal{B}_{1} \cup \mathcal{B}_{0}} \omega_{T,\ell}\lvert \beta_{\ell}\rvert\bigg],
\end{equation}
where \(\overline{g}_{T}(\alpha, \beta)=\frac{1}{T} \sum_{t=1}^{T} [
g_G(V_{G,t}, \alpha),
g_D(V_{D,t}, \alpha)-\beta]' \) represents the sample average of the moment conditions for both ``Good'' and ``Doubt'' instruments, $S_T$ denotes a symmetric weighting matrix, and $\lambda_T$ indicates a tuning parameter that calibrates the penalty applied to each slackness parameter $\beta_{\ell}$, encouraging sparsity in the selection of moment conditions. The penalty term $\sum_{\ell \in \mathcal{A} \cup \mathcal{B}_{1} \cup \mathcal{B}_{0}} \omega_{T,\ell}\lvert \beta_{\ell}\rvert$ plays a crucial role in determining which moments are retained in the final model, where \(\omega_{T,\ell}\) acts as an adaptive weight for each slackness parameter. This adaptive adjustment term, defined by \(\omega_{T,\ell}=\dot{\mu}_{T,\ell}^{r_{1}}\lvert \dot{\beta}_{T,\ell}\rvert^{-r_{2}}\), incorporates both empirical information and preliminary estimates to refine moment selection. The parameter $\dot{\mu}_{T,\ell}$, as further discussed below in Remark \ref{rew3.2}, quantifies the information content of moment $\ell \in {D}$, thereby emphasizing moments that contribute substantial information to the model. The term $\dot{\beta}_{T,\ell}$ represents a preliminary consistent estimator of ${\beta}_{0,\ell}=\mathbb{E}[g_D(V_{D,t,\ell}, \alpha_0)]$. By setting positive constants $r_1$ and $r_2$ ($r_2 \leq r_1$ ), we can control the sensitivity of the penalty, ensuring that the weight \(\omega_{T,\ell}\) diminishes for moments with negligible relevance, thereby reducing the likelihood of including irrelevant or redundant instruments in the model. By minimizing the objective function \eqref{eq14}, P-GMM achieves an optimal balance between bias and variance, leveraging the penalty to enforce a parsimonious model that reduces estimation error while retaining only the relevant moments. This approach is particularly beneficial in large-dimensional settings where the selection of valid and relevant\\ moments is computationally intensive.\looseness=-1

\begin{remark}
The penalized objective function in \eqref{eq14} is indeed a LASSO-type estimator, where each individual slackness parameter $\beta_{\ell}$ is penalized using its $L_1$-norm. This regularization technique encourages sparsity by shrinking irrelevant or weak moment conditions toward zero. In the context of GMM, this penalty structure allows for selective pruning of moments that do not substantially improve the efficiency of the estimator, as less informative or potentially invalid instruments are down-weighted or excluded from the estimation process. Theoretically, a moment $\ell$ will contribute to the final estimation if and only if $\lvert \beta_{\ell} \rvert>\lambda_T \omega_{T,\ell}/T$. The LASSO-type penalization thus refines the estimation by prioritizing moments with high informational value, directly influencing the asymptotic properties of the estimator.\looseness=-1
\end{remark}

\begin{remark}\label{rew3.2}
To effectively differentiate between valid and invalid moments, the preliminary estimator $\dot{\beta}_{T,\ell}$ plays a dominant role. For valid moments (Definition \ref{def1}), $\dot{\beta}_{T,\ell}$ is expected to exhibit a reduced magnitude, which results in a relatively high penalty within the objective function. Conversely, for invalid moments, $\dot{\beta}_{T,\ell}$ tends to be larger, leading to a smaller penalty. To further distinguish between relevant and irrelevant moments (Definition \ref{def2}), the empirical measure of information $\dot{\mu}_{T,\ell}$ is pivotal. For relevant moments, $\dot{\mu}_{T,\ell}$ is expected to be large, thereby driving the corresponding slackness parameter $\beta$ toward zero. For irrelevant moments, however, $\dot{\mu}_{T,\ell}$ will asymptotically approach zero, producing minimal shrinkage and allowing these moments to be naturally excluded from the final estimation. This dual mechanism ensures that both validity and relevance are taken into account, refining the P-GMM estimator by selectively incorporating only the moments that optimize estimation accuracy and efficiency.\looseness=-1
\end{remark}

\vspace{-0.11cm}

The parameter $\dot{\mu}_{T,\ell}$ in \(\omega_{T,\ell}\) is determined based on a relevance criterion given by
\vspace{-0.14cm}
\begin{equation}\label{eq15}
\dot{\mu}_{T, \ell}=\rho_{\max }\big(\hat{B}'_{G+\ell}\hat{S}^{-1}_{G+\ell} \hat{B}_{G+\ell}-\hat{B}'_{G} \hat{S}^{-1}_{G} \hat{B}_{G}\big), \ \ell \in {D}=\mathcal{A} \cup \mathcal{B}_{1} \cup \mathcal{B}_{0},
\vspace{-0.14cm}
\end{equation}
in which $\hat{B}_{G}$ and $\hat{S}_{G}$ correspond to estimators derived analogously to $\hat{B}_{T}$ and $\hat{S}_{T}$ as defined in Section \ref{sec:EKT}. The term $\rho_{\max}$ captures the maximum increase in information content gained by adding moment condition $\ell$,  highlighting the additional contribution of the moment $\ell$ to estimator efficiency. Specifically, $(\hat{B}'_{G} \hat{S}^{-1}_{G} \hat{B}_{G})^{-1}$ represents the asymptotic variance of the optimal estimator for $\alpha_0$ prior to adding the moment, while $(\hat{B}'_{G+\ell}\hat{S}^{-1}_{G+\ell} \hat{B}_{G+\ell})^{-1}$ reflects the asymptotic variance once the moment condition $\ell$ is included. The criterion thus assesses the relevance of the moment $\ell$ in terms of its informational gain relative to previously selected moments in $G$, incorporating only those conditions that significantly enha-\\nce estimation efficiency. We then have the following lemma demonstrating the efficiency gain.\looseness=-1

\begin{lemma}\label{lemma1}
Suppose $\hat{S}_{G}$ and $\hat{S}_{G+\ell}$ are invertible. Then, $\hat{B}'_{G+\ell}\hat{S}^{-1}_{G+\ell} \hat{B}_{G+\ell} \geq \hat{B}'_{G} \hat{S}^{-1}_{G} \hat{B}_{G}$ for all $\ell \in {D}$.
\end{lemma}

Lemma \ref{lemma1} establishes that incorporating a valid moment $\ell$ does not diminish the efficiency of the P-GMM post-selection estimator. This property ensures that the P-GMM procedure is capable of retaining only those moment conditions that substantively contribute to estimator precision, systematically refining the estimator by integrating the most informative moments. This mechanism is crucial in high-dimensional settings, where only a subset of moments enhances estimator accuracy and efficiency, thus\\ optimizing the asymptotic behavior of the estimator in large samples.\looseness=-1

The initial estimates of the parameters $\dot{\alpha}_{T}$ and $\dot{\beta}_{T,\ell}$ can be obtained by setting $\lambda_T=0$ in \eqref{eq14}, essentially applying the unpenalized P-GMM criterion. Thereafter, the value of $\dot{\mu}_{T,\ell}$ could be obtained from \eqref{eq15}. Following \cite{CL15}, the tuning parameter $\lambda_T$ for the $\ell$th moment condition is\\ calculated as\looseness=-1
\vspace{0.05cm}
\begin{equation}
\hat{\lambda}_{T, \ell}=2\big\| \hat{S}_{T}^{-\frac{1}{2}}(\ell) \widehat{\Pi}_{T}\big\| d^{\frac{r_{2}}{4}} T^{-\frac{1}{2}-\frac{r_{2}}{4}},
\vspace{0.05cm}
\end{equation}
where $\| \cdot \|$ represents the Euclidean norm, $\hat{S}_{T}^{-1}(\ell)$ denotes the $\ell$th row of the matrix $\hat{S}_{T}^{-1}$, $\widehat{\Pi}_{T}$ is the estimator of \(\Pi_{T}=I_{d}-S_{T}^{-\frac{1}{2}} \Gamma_{\alpha}\left(\Gamma_{\alpha}^{\prime} S^{-1}_{T} \Gamma_{\alpha}\right)^{-1} \Gamma_{\alpha}^{\prime} S_{T}^{-\frac{1}{2}}\), and $\Gamma_{\alpha}$ indicates the first derivative of $\overline{g}_T(\alpha,\beta)$. The introduction of the $L_1$ penalty in \eqref{eq14} enables the P-GMM estimator to achieve selection consistency. This consistency is achieved by correctly distinguishing between valid, relevant, and irrelevant moment conditions. Specifically, in large samples, P-GMM asymptotically identifies non-zero parameters ${\beta}_{\ell}$ in sets $\mathcal{B}_0$ and $\mathcal{B}_1$ as nonzero, indicating their irrelevance or invalidity. In contrast, it consistently estimates parameters within $\mathcal{A}$ as zero, effectively excluding any invalid or irrelevant moments. Thus, the P-GMM approach achieves consistent moment selection, systematically incorporating all valid and relevant moment conditions while excluding those that do not contribute meaningfully to the estimation\\ of $\alpha$ within the EKT (2005) framework.\looseness=-1

\begin{theorem}\label{thm1}
\indent Let $\hat{\beta}_{\ell}$ denote the estimator of the slackness parameter indexed by $\ell$ from \eqref{eq14}. Under the model specifications and conditions established in previous sections, the following selection consis-\\tency results hold asymptotically\looseness=-1
\begin{equation*}
\operatorname{P}\left(\cup_{\ell \in \mathcal{B}_{1}}\left\{\hat{\beta}_{\ell}=0\right\}\right) \rightarrow 0 \ \text { as } T \rightarrow \infty \quad \text{(Invalid)},
\vspace{-0.05cm}
\end{equation*}
\begin{equation*}
\operatorname{P}\left(\cup_{\ell \in \mathcal{B}_{0}}\left\{\hat{\beta}_{\ell}=0\right\}\right) \rightarrow 0 \ \text { as } T \rightarrow \infty \quad \text{ (Irrelevant)},
\end{equation*}
\begin{equation*}
\operatorname{P}\left(\cap_{\ell \in \mathcal{A}}\left\{\hat{\beta}_{\ell}=0\right\}\right) \rightarrow 1 \ \text { as } T \rightarrow \infty \quad \text{(Valid and Relevant)}.
\end{equation*}
\end{theorem}

Theorem \ref{thm1} establishes the asymptotic behavior of the moment selection mechanism within the P-GMM estimator, demonstrating that invalid or irrelevant moments are asymptotically excluded while valid and relevant moments are retained to enhance estimator performance within the EKT (2005) framework. These results demonstrate that, with probability approaching one, the P-GMM procedure can reliably distinguish between valid, relevant moments and those that are either invalid or irrelevant. Consistent with established GMM theory \citep{H82, H05}, incorporating additional valid instruments leads to efficiency gains, as shown in Lemma \ref{lemma1}. As a result, the P-GMM approach not only identifies and selects all valid and relevant moment conditions but also optimizes the estimation of $\alpha$\\ through lower standard errors, thus improving estimator precision and robustness.\looseness=-1

From an asymptotic perspective, selecting only highly informative moments enhances the efficiency of the estimator
$\hat{\alpha}_0$ by focusing the information in $(B'S^{-1}B)^{-1}$ on the moments contributing most significantly. We can then achieve the following asymptotic normality with a reduced variance, aligning with EKT (2005) $Proposition$ 4 and leveraging the asymptotic properties of GMM.\looseness=-1

\begin{theorem}\label{thm2}
Under consistent moment selection in Theorem 1, the asymptotic distribution of the P-GMM post-selection estimator $\hat{\alpha}_0$ is given by\looseness=-1
\vspace{-0.15cm}
\begin{equation*}
\sqrt{T}\left( \hat{\alpha}_0-\alpha_0 \right) \stackrel{d}{\rightarrow} \mathcal{N}\left(0,(B_{\mathcal{A}}^{\prime} S_{\mathcal{A}}^{-1} B_{\mathcal{A}})^{-1}\right),
\vspace{-0.15cm}
\end{equation*}
where $B_{\mathcal{A}}=\mathbb{E}(\partial \bar{g}_T(\alpha_0,\beta_0)/\partial \alpha)$ for moments in ${\mathcal{A}}$ and $S_{\mathcal{A}}$ denotes the corresponding symmetric weighting matrix, ensuring only valid and relevant moments contribute to the limiting distribution.\looseness=-1
\end{theorem}

\vspace{-0.07cm}

Theorem \ref{thm2} demonstrates that the P-GMM post-selection estimator achieves efficiency comparable to an oracle GMM estimator based on all valid and relevant moments, as supported by the simulation results below. This oracle property highlights the potential of P-GMM estimation to enhance the power of rationality and symmetry tests by reducing the impact of uninformative instruments. Consequently, by systematically excluding moments that are either irrelevant or invalid, P-GMM effectively optimizes estimator precision, thereby bolstering the robustness and efficiency of the estimator in prac-\\tical applications where instrument selection poses a critical challenge.\looseness=-1

\begin{remark}\label{re3.3}
In our theoretical analysis, we follow the original setup of EKT (2005), treating the forecast \( f^*_{t+h} = \theta^{*\prime} W_t \) as a primitive object that is directly generated by the forecaster according to an optimal decision rule. We do not account for the estimation error that may arise when forecasts are constructed using recursive, rolling, or fixed-window schemes methods. This approach enables us to focus on the identification of the forecaster's preference parameter $\alpha$, based on valid moment conditions. As such, the validity of our asymptotic results relies on the assumption that forecasts are conditionally exo-\\genous and fixed with respect to the instrument set used for moment selection.\looseness=-1
\end{remark}

\begin{remark}
The distinction between ``good'' and ``doubt'' instruments is central to the P-GMM procedure. In practice, we recommend including in the ``good'' set \( V_{G,t} \) those instruments that satisfy at least one of the following criteria: (i) {economically justified}, i.e., variables that are commonly employed across forecasting models, such as lagged dependent variables; (ii) {statistically indispensable}, i.e., instruments whose exclusion would compromise the identification of the forecaster's preference parame-\looseness=-1

\noindent ter \( \alpha \); and (iii) {historically robust}, i.e., instruments that have demonstrated consistent validity in prior studies, such as constant terms or benchmark control variables. For example, in both our simulations and empirical application, we include the constant and the lagged dependent variable in \( V_{G,t} \), consistent with standard practice in forecast rationality testing. We recommend starting with a parsimonious set of core instruments supported by strong theoretical justification, while assigning all other instruments, including nonlinear transformations, higher-order terms, and novel predictors, to the ``doubt'' set \( V_{D,t} \),\\ which is then subject to penalized selection.\looseness=-1
\end{remark}

\section{Monte Carlo Simulations} \label{sec:numerical}
\noindent We conduct simulation experiments to evaluate the finite sample properties of the proposed P-GMM estimation within the EKT (2005) framework, analyzing cases of both linear and nonlinear dependencies between the forecast user's information set and forecast error. For both scenarios, we treat the constant instrument as part of the valid and relevant set $V_{G,t}$ defined in Section \ref{sec:selection}. Throughout this section, we employ the identity matrix as the weighting matrix in the GMM optimization to avoid additional parameter estimation, with the forecast horizon set to $h=1$. We explore the behaviour of the estimator of $\alpha$ and investigate the characteristics of the $J$-test statistics to underscore the impact of instrum-\\ental variable selection on estimator performance.\looseness=-1

\indent \textbf{DGP 1: Linear Case} \ To investigate the efficacy of the proposed P-GMM approach in a high-dimensional setting, we consider a scenario with $d=30$ instruments in $V_t$. We generate the random\\ samples from the following data generating process (DGP)\looseness=-1
\vspace{-0.14cm}
\begin{equation}
Y_{t+1}=\theta_0+W_{1,t}\theta_1+W_{2,t}\theta_2+U_t,
\vspace{-0.14cm}
\end{equation}
where $Y_{t+1}$ represents the dependent variable, $W_{1,t}$ and $W_{2,t}$ are covariates, $U_t$ denotes the error term, and the parameter values are set to $\theta_0=1$, $\theta_1=0.5$ (strong identification) or 0.07 (weak identification), and $ \theta_2=0.4$, capturing a baseline configuration with significant contributions from each predictor. In this simulation, the goal is to differentiate the impact of valid, invalid, and irrelevant instruments on the estimation and the moment selection procedure. To achieve this, the simulated instrument set includes both a valid but irrelevant instrument vector $W_{v,t}$ and an invalid instrument vector $W_{inv,t}$. The\\ samples are generated through the following multivariate normal distribution\looseness=-1
\begin{equation}
\begin{aligned}
& \quad \quad \quad \quad \quad \left( \begin{array}{c}{W_ {1,t}} \\ {W_{2,t}} \\ {W_{v,t}} \\ {W^*_{v,t}} \\ {U_{t}}\end{array}\right)  \sim N \left( 0, \left( \begin{array}{ccccccccc} {1} & {0.2} & {0} & {0}  & {0} \\ {0.2} & {1} & {0}  & {0} & {0} \\ {0} & {0} & {I_{14}} & {0}  & {0}  \\{0} & {0} & {0} & {I_{14}}  & {0} \\{0} & {0} & {0} & {0}  & {1} \end{array} \right) \right), \\[4pt]
& W_{inv,t}(\ell)=W^*_{v,t}(\ell)+c_\ell \times U_t, \ \text{and} \ c_{\ell}=c_{0}+\frac{(\ell-1)\left(\overline{c}-c_{0}\right)}{d / 2-1}, \ \ell=1, \dots, d-k_0
\end{aligned}
\end{equation}
with $k_0=3$ for $V_{G,t}=(1,W_{1,t},W_{2,t})'$, where $ W_{inv,t}(\ell)$ and $W^*_{v,t}(\ell)$ are the $\ell$th instruments in the simulated data. The formulation of $W_{inv,t}(\ell)$ incorporates a linear dependence on the error term $U_t$ through coefficients $c_{\ell}$ that vary across the set, enabling us to systematically examine the impact of invalid instruments of differing strengths on the estimator. Since $W_{inv,t}(\ell)$ is constructed to be correlated with $U_t$, it does not meet the criteria for validity within the set of instruments used for estimating and test-\\ing forecast rationality.\looseness=-1

\indent \textbf{DGP 2: Nonlinear Case} \ In this simulation setting, we explore a nonlinear DGP with a relati-\\vely high dimensionality, specifying $d=31$ instruments in $V_t$. The model is specified as follows\looseness=-1
\vspace{-0.1cm}
\begin{equation}
Y_{t+1}=\theta_0+W^2_{1,t}\theta_1+(W_{1,t}\times W_{2,t})\theta_2+\exp(W_{3,t})\theta_3+U_t,
\vspace{-0.1cm}
\end{equation}
where $Y_{t+1}$ represents the dependent variable, and $W_{1,t}$, $W_{2,t}$, and $W_{3,t}$ are predictor variables. Each predictor is transformed through nonlinear functions to introduce diverse structural relationships within the model. The parameter values are set to $\theta_0=1$, $\theta_1=0.5$ or 0.07, $ \theta_2=0.5$, and $\theta_3=0.4$, reflecting a scenario with varying weights on each component. Similar to DGP 1, the instrument set is augmented to include both a valid but irrelevant instrument vector $W_{v,t}$ and an invalid vector $W_{inv,t}$.\\ The samples are generated according to the following multivariate normal distribution\looseness=-1
\begin{equation}
\begin{aligned}
& \quad \quad \quad \quad \left( \begin{array}{c}{W_ {1,t}} \\ {W_{2,t}} \\ {W_{3,t}} \\ {W_{v,t}} \\ {W^*_{v,t}} \\ {U_{t}}\end{array}\right)  \sim N \left( 0, \left( \begin{array}{ccccccccc} {4} & {0.2} & {0.04}& {0}  & {0}  & {0} \\ {0.2} & {10} & {0.2} & {0} & {0} & {0} \\ {0.04} & {0.2} & {8} & {0} & {0} & {0} \\ {0} & {0} & {0} & {I_{14}} & {0}  & {0}  \\{0} & {0} & {0} & {0} & {I_{14}}  & {0} \\{0} & {0} & {0} & {0} & {0}  & {1} \end{array} \right) \right), \\[4pt]
&
W_{inv,t}(\ell)=W^*_{v,t}(\ell)+c_\ell \times U_t, \ \text{and} \ c_{\ell}=c_{0}+\frac{(\ell-1)\left(\overline{c}-c_{0}\right)}{d / 2-1}, \ \ell=1, \dots, d-k_0
\end{aligned}
\end{equation}
with $k_0=4$ for the set of valid instruments $V_{G,t}=(1,W^2_{1,t},W_{1,t}W_{2,t},\exp(W_{3,t}))'$. By introducing correlations between $W_{inv,t}(\ell)$ and $U_t$ through varying coefficients $c_{\ell}$, we systematically incorporate deg-\\rees of invalidity within the instrument set. \looseness=-1

In both DGPs, the initial parameters are set to $c_0=0.5$ and an upper bound $\overline{c}=2.4$, with user-selected constants $r_1=3$ and $r_2=2$ chosen in line with \cite{CL15}. We investigate two sample sizes, $T=200$ and $T=2000$, each evaluated over 5000 simulation repetitions to ensure robustness. In all simulation experiments, each dataset is split such that 50\% of the observations are used by the forecast user to estimate the forecasting model parameters \( \theta \). The estimated coefficients \( \hat{\theta} \) are then used to generate one-step-ahead forecasts for the remaining 50\% of the sample. This fixed-window design preserves parameter estimation error in the forecasts, allowing us to assess the finite sample performance of the proposed P-GMM estimator under realistic conditions. As illustrated in Remark \ref{re3.3}, the induced estimation uncertainty is not incorporated in the theoretical analysis, which treats forecasts as primitives, but is accounted for in the simulations to evaluate the robustness of the inference procedures. The forecasts are generated by estimating $\hat{\theta}$ for a fixed loss shape parameter $p_0=2$, given a selected subset of variables from the forecast user's information set and the forecaster's known asymmetric preference parameter $\alpha_0$, where $\hat{\theta}$ represents the estimated parameter vector within each respective DGP framework. Note that it is reasonable to fix the loss shape parameter $p$ since \cite{KO12} have demonstrated that different values of $p$ yield consistent estimates for the asymmetric parameter $\alpha$. The asymmetric preference parameter is varied across the values $\alpha_0 \in \{0.2,0.4,0.5,0.6,$ $0.8\}$ to encompass a spectrum of forecasting behaviors, from highly asymmetric preferences through values approaching symmetry to the fully symmetric case. Subsequently, we estimate different values of $\alpha$ based on the resultant one-step-ahead forecast errors using the P-GMM estimation approach, enabling us to assess the consistency and efficiency of $\alpha$ estimates across differing levels of\\ asymmetry in user preferences.\looseness=-1

\vspace{0.1cm}

{\captionof{table}{Performance of P-GMM Estimation (DGP 1)} \label{tab:tab1}
\vspace{-0.1cm}
\begin{table}[h]
\renewcommand\arraystretch{1.31}
\begin{center}
\resizebox{10cm}{!}{
\begin{tabular}{ccccccc p{1cm}}
\hline\hline
\multirow{2}{*}{$\alpha$} & \multicolumn{1}{c}{$T=200$} & & &  \multicolumn{1}{c}{$T=2000$} \\ \cline{2-7} 
                   & P(VR)  & P(VR+)  & P(INV)  & P(VR)   & P(VR+)   & P(INV)  \\ \hline
\multicolumn{4}{c}{{\underline{Strong Identification ($\theta_1=0.5$)}}}\\                  
0.2               & 0.676  & 0.138   & 0       & 0.958   & 0.032    & 0       \\
0.4               & 0.629  & 0.141   & 0       & 0.961   & 0.024    & 0      \\
0.5               & 0.683  & 0.109   & 0       & 0.967   & 0.021    & 0       \\
0.6               & 0.647  & 0.193   & 0       & 0.962   & 0.016    & 0       \\
0.8               & 0.684  & 0.112   & 0       & 0.953   & 0.027    & 0       \\ \hline
\multicolumn{4}{c}{{\underline{Weak Identification ($\theta_1=0.07$)}}}\\                  
0.2               & 0.104  & 0.872   & 0.016       & 0.892   & 0.108    & 0       \\
0.4               & 0.117  & 0.855   & 0.020       & 0.906   & 0.091    & 0      \\
0.5               & 0.133  & 0.850   & 0.016       & 0.874   & 0.120    & 0       \\
0.6               & 0.105  & 0.871   & 0.023       & 0.892  & 0.100    & 0       \\
0.8               & 0.124  & 0.858   & 0.018       & 0.898   & 0.102    & 0       \\ \hline\hline
\end{tabular}}
\vspace{-0.1cm}
  \end{center}
\small{{\bf{\textit{Note}}}: In 5000 simulation repetitions, ``P(VR)'' represents the probability that P-GMM estimation selects all valid and relevant moments; ``P(VR+)'' indicates the probability that P-GMM estimation selects all valid and relevant moments along with some irrelevant ones; and ``P(INV)'' denotes the probability that P-GMM esti-\\mation selects invalid moment conditions.\looseness=-1}
\end{table}}

The performance of P-GMM estimation for DGP 1 with various values of the asymmetric parameter is presented in Table \ref{tab:tab1}. For the strong identification scenario, we observe that for each value of $\alpha$, the probability of selecting invalid moment conditions is consistently 0, indicating that P-GMM estimation effectively selects only the valid moment conditions. With a smaller sample size of $T=200$, the probability of P-GMM estimation capturing all valid and relevant moments (specifically, $W_1$ and $W_2$) in the context of EKT (2005) is approximately 0.65, while the probability of selecting all valid and relevant moments plus a few irrelevant ones (from $W_{inv}$) is around 0.15. Consequently, in the EKT (2005) framework, P-GMM estimation identifies the complete set of valid and relevant moments with a probability close to 0.8. As the sample size increases to 2000, the probability of correctly identifying all valid and relevant moment conditions improves to approximately 0.95, and the likelihood of incorporating irrelevant moments remains close to 0. These results suggest strong finite sample performance of P-GMM estimation. In the context of EKT (2005), we further assess P-GMM estimation's efficacy under weak identification conditions. For smaller sample sizes, the probability of selecting only valid and relevant moments is reduced, yet the probability of capturing all valid and relevant moments approaches 0.85. With a larger sample size of 2000, the probability of P-GMM estimation mistakenly including\\ invalid moment conditions remains exactly 0.\looseness=-1

\vspace{0.1cm}

{\captionof{table}{Performance of P-GMM Estimation (DGP 2)} \label{tab:tab2}
\vspace{-0.1cm}
\begin{table}[h]
\renewcommand\arraystretch{1.31}
\begin{center}
\resizebox{10cm}{!}{
\begin{tabular}{ccccccc p{1cm}}
\hline\hline
\multirow{2}{*}{$\alpha$} & \multicolumn{1}{c}{$T=200$} & & &  \multicolumn{1}{c}{$T=2000$} \\ \cline{2-7} 
                   & P(VR)  & P(VR+)  & P(INV)  & P(VR)   & P(VR+)   & P(INV)  \\ \hline
\multicolumn{4}{c}{{\underline{Strong Identification ($\theta_1=0.5$)}}}\\                  
0.2               & 0.633  & 0.174   & 0       & 0.932   & 0.063    & 0       \\
0.4               & 0.617  & 0.178   & 0       & 0.947   & 0.050    & 0      \\
0.5               & 0.685  & 0.182   & 0       & 0.953   & 0.039    & 0       \\
0.6               & 0.635  & 0.195   & 0       & 0.949   & 0.048    & 0       \\
0.8               & 0.663  & 0.168   & 0       & 0.947   & 0.044    & 0       \\ \hline
\multicolumn{4}{c}{{\underline{Weak Identification ($\theta_1=0.07$)}}}\\                  
0.2               & 0.125  & 0.820   & 0.029       & 0.909   & 0.083    & 0       \\
0.4               & 0.113  & 0.864   & 0.021       & 0.910   & 0.086    & 0      \\
0.5               & 0.142  & 0.831   & 0.023       & 0.886   & 0.084    & 0       \\
0.6               & 0.126  & 0.835   & 0.035       & 0.908  & 0.089    & 0       \\
0.8               & 0.137  & 0.826   & 0.024       & 0.896   & 0.091    & 0       \\ \hline\hline
\end{tabular}}
  \end{center}
\end{table}}

\vspace{-0.5cm}

The results for DGP 2 in Table \ref{tab:tab2} reveal similar performance trends in P-GMM estimation as seen in DGP 1, with slight differences in probabilities due to the nonlinear structure of DGP 2. In the strong identification case with a smaller sample size of $T=200$, the probability of selecting all valid and relevant moments is slightly lower than in DGP 1, ranging from 0.617 to 0.685 across different values of $\alpha$, reflecting the added complexity from nonlinear dependence. The probability of selecting all valid and relevant moments plus some irrelevant ones remains similar to DGP 1, around 0.17 to 0.19, while the probability of selecting invalid moments is zero, confirming P-GMM's effectiveness in excluding invalid moments. When the sample size increases to $T=2000$, the probability of selecting only valid and relevant moments increases to approximately 0.95, with a slight reduction in selecting additional irrelevant moments. In the weak identification case, P-GMM's performance parallels DGP 1, though with marginally higher probabilities of selecting irrelevant moments due to the nonlinear effects. Overall, P-GMM estimation demonstrates reliable finite sample performance, particularly in excluding invalid moments while retaining all valid and relevant moments. The nonlinear structure slightly impacts pre-\\cision in smaller samples, yet the estimator remains robust and effective as the sample size increases.\looseness=-1

In Tables \ref{tab:tab3} and \ref{tab:tab4}, we present the finite sample properties of the estimator $\alpha$ for DGP 1 and DGP 2, comparing five different scenarios: (a) P-GMM estimation; (b) estimation using all valid and relevant moment conditions; (c) estimation using all moment conditions; (d) estimation with a constant instrument; and (e) estimation with a constant instrument plus instrument $W_1$. These varying instrument sets include both omitted and irrelevant information, providing a comprehensive comparison across scenarios with different levels of instrument relevance and validity. All estimates are calculated based on\looseness=-1

\clearpage

\begin{footnotesize}
\centering
\begin{sidewaystable}
\scriptsize
\captionof{table}{Estimation and Testing Results---DGP 1} \label{tab:tab3}
\vspace{-0.6cm}
\renewcommand\arraystretch{1.07}
\begin{center}
\resizebox{20.6cm}{!}{
\begin{tabular}{cccccccccccc p{1.1cm}}
\hline\hline
\multirow{4}{*}{} & \multicolumn{1}{c}{$T=200$}            & & & & & \multicolumn{1}{c}{$T=2000$}          \\ \cline{2-11} 
                  & P-GMM    & VR    & All    &CON   &CON+$W_1$ & P-GMM    & VR    & All     &CON  &CON+$W_1$  \\   \hline
$\alpha=0.2$               & 0.2075 & 0.2001 & 0.0351 &0.2001 &0.2000 & 0.2000 & 0.2000  & 0.0212 &0.2000 &0.2000\\ 
se                & 0.02107 & 0.02106 & 0.02123 &0.02204 &0.02194 & 0.00509 & 0.00507  & 0.00539 &0.00515 &0.00511\\ 
$J$(0.5)-Stat    &18.2595 &14.6290   &- &6.5038 &11.4496 &16.8757 &13.6572 &- &7.4624 &11.5735\\
$RF$ &0.0030 &0.0026 &- &0.0116 &0.0036 &0.0024 &0.0014 &- &0.0066 &0.0036\\
$J$-Stat &1.5834 &0.0402 &- &- &0.2725  &1.3424 &0.0310  &- &- &0.1578\\
$RF$ &0.8122 &0.8414 &- &- &0.6020 &0.8544 &0.8606 &- &- &0.6916\\   \hline
$\alpha=0.4$               & 0.4030 & 0.4000 & 0.2803 &0.4000 &0.4000 & 0.4000 & 0.4000 & 0.1865 &0.4000 &0.4000\\ 
se                & 0.02701 & 0.02699 & 0.02804 &0.02711 &0.02707 & 0.00516 & 0.00515  & 0.00581 &0.00619 &0.00618\\ 
$J$(0.5)-Stat &18.5628 &12.6832 &- &5.7825 &8.6394 &20.5735 &13.0274  &- &5.8582 &9.4294\\
$RF$ & 0.0102 &0.0024 &- &0.0166 &0.0140 &0.0026 &0.0020 &- &0.0160 &0.0094\\
$J$-Stat &2.6042 &0.0124 &- &- &0.6526 &2.2872 &0.0118 &- &- &0.4892\\
$RF$ &0.8574 &0.9120 &- &- &0.4198 &0.8920 &0.9142 &- &- &0.4848\\   \hline
$\alpha=0.5$              & 0.4999 & 0.5000 & 0.2741 & 0.4999 &0.4999 & 0.5000 & 0.4999 & 0.3029 &0.4999 &0.4999\\ 
se                & 0.02789 & 0.02781 & 0.02807 &0.02821 &0.02803 & 0.00504 & 0.00503  & 0.00544 &0.00539 &0.00520\\ 
$J$(0.5)-Stat &5.3682 &0.2402 &- &0.1356 &0.4982 &2.3835 &0.1801 &- &0.1287 &0.4026\\
$RF$ &0.7182 &0.8874  &- &0.7134 &0.7802 &0.7946 &0.9146 &- &0.7204 &0.8184\\
$J$-Stat &5.3682 &0.2402 &- &- &0.4982 &2.3835 &0.1801 &- &- &0.4026\\
$RF$  &0.6158 &0.6248 &- &- &0.4810 &0.6662 &0.6720 &- &- &0.5258\\   \hline
$\alpha=0.6$               & 0.5990 & 0.6009  & 0.2303 &0.5999 &0.5999 & 0.6000 & 0.5999 & 0.3135 &0.5999 &0.6000\\ 
se                & 0.02689 & 0.02688  & 0.02821 &0.02735 &0.02701 & 0.00581 & 0.00579  & 0.00596 &0.00609 &0.00599 \\ 
$J$(0.5)-Stat &17.5738 &13.8264 &- &5.4394 &9.5727 &16.7295 &14.8463 &- &5.9592 &10.4862\\
$RF$ &0.0148 &0.0016 &- &0.0204 &0.0090 &0.0110 &0.0012 &- &0.0152 &0.0060\\
$J$-Stat &2.6724 &0.0214 &- &- &0.6482 &1.8913 &0.0186 &- &- &0.5402\\
$RF$ &0.8494 &0.8844 &- &- &0.4214 &0.8646 &0.8922 &- &- &0.4630\\   \hline
$\alpha=0.8$                & 0.7993 & 0.8000 & 0.4766 &0.8000 &0.8000 & 0.8000 & 0.8000 & 0.3272 &0.8000 &0.8000 \\ 
se                & 0.02735 & 0.02705 & 0.04355 &0.02984 &0.02861 & 0.00599 & 0.00582 & 0.00814 &0.00645 &0.00617\\ 
$J$(0.5)-Stat & 16.8352 &13.6572 &- &6.7624 &5.7624 &19.4682 &13.9253 &- &6.8410 &5.6842\\
$RF$ &0.0106 &0.0018&- &0.0170 &0.0568 &0.0074 &0.0016 &- &0.0096 &0.0590\\
$J$-Stat &1.7826 &0.1245 &- &- &0.6873 &2.2765 &0.1189 &- &- &0.5924\\
$RF$ &0.8790 &0.7248 &- &- &0.4078 &0.8932 &0.7308 &- &- &0.4422\\   \hline\hline
\end{tabular}}
\vspace{-0.2cm}
 \end{center}
 \footnotesize{{\bf{\textit{Note}}}: The first row related to $\alpha$ reports the average estimates over 5000 repetitions, with ``se'' representing the average standard error calculated from these 5000 estimates. ``VR'' signifies estimation using all valid and relevant moments, while ``All'' indicates estimation based on all moments. ``CON'' means estimation using only a constant as an instrument, and ``CON+$W_1$'' refers to estimation using a constant plus the instrument $W_1$. For each $\alpha$ value, the first $RF$ (rejection frequency)\\ corresponds to the $J$(0.5)-statistic and the second $RF$ corresponds to the $J$-statistic, with both values averaged over 5000 repetitions.\looseness=-1}
\end{sidewaystable}
\end{footnotesize}

\clearpage

\begin{footnotesize}
\centering
\begin{sidewaystable}
\scriptsize
\captionof{table}{Estimation and Testing Results---DGP 2} \label{tab:tab4}
\vspace{-0.6cm}
\renewcommand\arraystretch{1.07}
\begin{center}
\resizebox{20.6cm}{!}{
\begin{tabular}{cccccccccccc p{1.1cm}}
\hline\hline
\multirow{4}{*}{} & \multicolumn{1}{c}{$T=200$}            & & & & & \multicolumn{1}{c}{$T=2000$}     \\ \cline{2-11} 
                  & P-GMM    & VR    & All    &CON &CON+$W^2_1$ & P-GMM    & VR    & All     &CON  &CON+$W^2_1$  \\   \hline
$\alpha=0.2$               & 0.1995 & 0.1992 & 0.1027 &0.2109 &0.2093  &0.1999 &0.2000 &0.0907 & 0.1898 &0.2068 \\ 
se                & 0.02665 & 0.02194 & 0.02443 &0.02861 &0.02651 &0.00509 &0.00507 &0.00546 &0.00519 &0.00513 \\ 
$J$(0.5)-Stat    &12.7653 &17.4673   &- &4.2579 &8.2461  &10.5372 &9.4751 &- &5.0283 &8.8562\\
$RF$ &0.0260 &0.0020 &- &0.0396 &0.0166 & 0.0222 &0.0018 &- & 0.0308 &0.0192\\
$J$-Stat &2.4625 &1.3795 &- &- &0.5726 & 2.1780 &1.1677 &- &- &0.4193\\
$RF$ &0.6520 &0.7110 &- &- &0.4498 &0.7292 &0.7732 &- &- &0.4810 \\   \hline
$\alpha=0.4$  &0.4017 &0.4000 &0.1952 &0.3806 &0.3893 &0.4000 &0.4000 & 0.1729 &0.3973 &0.3999        \\ 
se    &0.02845 &0.02833 &0.04228 &0.02861 &0.02860   &0.00513 &0.00512 &0.00596 &0.00608 &0.00610         \\ 
$J$(0.5)-Stat & 16.5234 &16.3757 &- &4.3712 &8.6584 &18.5783 &17.9762 &- &5.0723 &9.5822\\
$RF$ & 0.0212  &0.0030 &- &0.0370 &0.0136 &0.0202 &0.0024 &- &0.0366 & 0.0178 \\
$J$-Stat & 4.9872  &1.6320 &- &- &0.9537 & 4.4577 & 1.4288 & - &- &0.7205 \\
$RF$ & 0.5462 &0.6528 &- &- &0.3294  &0.6124 &0.7254 &- &- &0.3826  \\   \hline
$\alpha=0.5$   &0.4987 &0.4997 &0.3318 &0.4972 &0.4972  &0.5000 &0.5000 &0.3168 &0.4894 & 0.4992       \\ 
se  &0.02053 &0.19277 &0.02148 &0.02099 &0.02136    &0.00503 &0.00501 &0.00553 &0.00530 &0.00529        \\ 
$J$(0.5)-Stat &3.8934 &1.0028 &- &0.1538 &0.4016  &2.6780 &0.8207 &- &0.1399 &0.3628 \\
$RF$  &0.8672 &0.9100 &- &0.6970 &0.8188  &0.9034 &0.9190 &- &0.7376 &0.8534 \\
$J$-Stat &3.8934 &1.0028 &- &- &0.4016 &2.7793 &0.6638 &- &- &0.3827 \\
$RF$ &0.7924 &0.8010 &- &- &0.5268 &0.8276 &0.8540 &- &- &0.5966 \\   \hline
$\alpha=0.6$  & 0.6002  &0.5994 &0.0184 &0.5793 &0.5825  &0.6000 &0.5999 &0.0347 &0.5797 &0.5803     \\ 
se  &0.02801 &0.02793 &0.04427 &0.02886 &0.02932 &0.00593 &0.00613 &0.00693 &0.00679 &0.00640           \\ 
$J$(0.5)-Stat &20.7642  &18.3573 &- &5.7263 &7.4657 &18.7682 &16.0324 &- &6.1836 &8.3029 \\
$RF$ &0.0026 &0.0018 &- &0.0174 &0.0246 &0.0024 &0.0016 &- &0.0150 &0.0224  \\
$J$-Stat &2.9242 &1.0937 &- &- &0.3538  & 2.4628 &1.0437 &- &- &0.3139 \\
$RF$ &0.7122 &0.7790 &- &- &0.5524 &0.7632 & 0.8210 &- &- &0.5726 \\   \hline
$\alpha=0.8$ &0.7993 &0.8000 &0.2952 &0.7903 &0.7922  &0.8000 &0.8000 &0.1758 &0.7993 &0.7999             \\ 
se   &0.01963 &0.01852 &0.04896 &0.02086 &0.02114 &0.00609 &0.00581 &0.00830 &0.00776 &0.00783          \\ 
$J$(0.5)-Stat &18.1683 &17.6478 &- &6.4548 &5.4762  &20.9632 &18.0027 &- &6.9273 &5.8829 \\
$RF$ &0.0116 &0.0018 &- &0.0116 &0.0652 &0.0932 &0.0014 &- &0.0096 &0.0616 \\
$J$-Stat &2.8535 &0.5972 &- &- &0.2923 &3.1762 &0.5128 &- &- &0.2380 \\
$RF$ &0.8276 &0.8978 &- &- &0.5894 &0.8536 &0.9308 &- &- &0.6224\\   \hline\hline
\end{tabular}}
\vspace{-0.2cm}
 \end{center}
 \footnotesize{{\bf{\textit{Note}}}: The notations are consistent with those in Table \ref{tab:tab3}, except that ``CON+$W^2_1$'' refers to estimation using a constant instrument combined with $W^2_1$.\looseness=-1}
\end{sidewaystable}
\end{footnotesize}

\clearpage

\noindent 5000 simulation repetitions to ensure statistical reliability, with values averaged across these runs. For each repetition, we compute the $J(0.5)$-statistic and $J$-statistic based on \eqref{eq11} and \eqref{eq12}. We also report empirical rejection frequencies of the $J(0.5)$-statistic and $J$-statistic at the 5\% significance level across 5000 simulation repetitions. These frequencies reflect the proportion of simulations in which the null hypothesis is rejected, thereby serving as an informative functional summary of the $p$-value distrib-\\ution under the alternative hypothesis.\looseness=-1

Table \ref{tab:tab3} provides a detailed comparison of the finite sample properties of the P-GMM estimator, showcasing its consistency and reduced standard error relative to other estimation approaches, with the exception of scenario (b), which serves as the oracle estimator, representing an ideal and theoretically attainable model. The precision of the P-GMM estimator for the asymmetry parameter $\alpha$ improves substantially with an increasing sample size, reinforcing its robustness under larger samples. In scenario (c), where estimation includes all available moment conditions, the estimator exhibits bias, shifting away from the true parameter value due to the presence of invalid instruments. This bias remains as the sample size grows, indicating that simply adding observations cannot mitigate the influence of invalid moments. Such biases can inflate type I error, raising the probability of erroneously rejecting valid null hypotheses in hypothesis testing. Given these limitations, we exclude the $J$-test statistics for scenario (c) in subsequent analyses of forecast rationality, as the inherent biases compromise the reliability of the test. This outcome illustrates the necessity of selective instrument inclusion when estimating and testing forecast rationality under flexible loss functions, as arbitrary instrument selection can introduce substantial biases. These findings therefore emphasize the critical role of rigorous moment selection within the EKT (2005) framework. The results for scenario (d), where only a constant instrument is included, show that the estimator remains consistent despite limited information, in line with theoretical expectations given the validity and relevance of a constant instrument. However, this approach lacks efficiency, as indicated by larger standard errors and reduced statistical power in comparison to scenarios (a) and (b). Notably, the P-GMM post-selection estimator in scenario (a) performs comparably to the oracle estimator in scenario (b). This alignment is theoretically anticipated, as P-GMM post-selection is designed to retain only the most informative and relevant moments, thereby improving estimator efficiency. This outcome highlights the effectiveness of P-GMM in refining estimator precision by prioritizing relevance and validity within the selected instrument set.\looseness=-1

In evaluating the tests conducted after P-GMM moment selection, we find that the rejection frequencies from the rationality and symmetry tests outperform all scenarios except for the oracle case (b). Across all values of $\alpha$, the rejection frequencies of the $J$-test statistics using P-GMM estimation exceed those of other estimation approaches, apart from the oracle estimator. For instance, at $\alpha = 0.2$, the P-GMM approach yields a rejection frequency of 0.8122 for the $J$-test statistic in a sample of size $T=200$, rising to 0.8544 as $T$ increases to 2000. This outcome demonstrates P-GMM's effectiveness in enhancing the power of the rationality test by selectively incorporating only valid and relevant moments. Specifically, for testing loss symmetry at $\alpha = 0.5$, the rejection frequency of the $J(0.5)$-test under P-GMM is 0.7182 at $T=200$ and improves to 0.7946 with $T=2000$, closely aligning with the oracle's superior results. In contrast, scenarios such as estimation using a constant instrument yield lower rejection frequencies (0.7134 at $T=200$ and 0.7204 at $T=2000$), reflecting lower sensitivity and effectiveness in capturing symmetry deviations. This result underscores that P-GMM can bolster the power of the symmetry test by judiciously choosing moments that best align with the structure of the model. For other asymmetry parameter values ($\alpha = 0.2$, 0.4, 0.6, and 0.8), the rejection frequencies of the $J(0.5)$-test statistics using P-GMM estimation are slightly higher than the oracle scenario but still outperform other cases. For example, at $\alpha = 0.8$, the rejection frequency for the $J(0.5)$-statistic is 0.0106 with $T=200$ and 0.0074 with $T=2000$, again outperforming scenarios involving all moments or less selective instrument sets. This outcome suggests that by reducing the impact of uninformative instruments, P-GMM strengthens the test's sensitivity to symmetry deviations, effectively capturing more precise estimates of the asymmetry parameter. Additionally, as the sample size increases, the power of the $J$-test statistics shows further improvement, validating P-GMM's capacity to achieve both theoretical rigor and practical efficacy in managing instrument selection. All of these highlight P-GMM's potential to deliver the theoretical advantages and computational efficiency needed to optimally filter ins-\\trumental variables for robust forecast rationality estimation and testing.\looseness=-1

\indent Table \ref{tab:tab4} represents the finite sample properties of the asymmetry parameter $\alpha$ for DGP 2, where we observe trends and patterns similar to those noted for DGP 1. As shown, P-GMM estimation demonstrates consistently robust performance across all values of $\alpha$, with estimates closely aligning to the true parameter values and lower standard errors than other scenarios. This finding highlights the effectiveness of P-GMM in selecting relevant instruments, even under nonlinear dependencies present in DGP 2. In the scenario utilizing only a constant as an instrument, the estimates of $\alpha$ show reasonable consistency, though the standard errors are relatively higher, reflecting a trade-off in precision due to limited instrument information. When the instrument set includes both the constant and $W^2_1$, the precision of $\alpha$ estimates improves slightly, yet remains inferior to P-GMM estimation. For cases incorporating all available moment conditions, we observe higher bias and standard errors, particularly in smaller samples. This trend, consistent with DGP 1, underscores the detrimental effect of including invalid instruments, which induces bias that cannot be mitigated by sample size alone. Such biases also inflate type I error, as indicated by the lower rejection frequencies for the $J(0.5)$- and $J$-statistics in these cases. Thus, including all moments risks over-rejecting the null hypothesis of forecast rationality and symmetry. Analyzing the rejection frequencies from rationality and symmetry tests, P-GMM estimation yields the highest rejection frequencies across most $\alpha$ values compared to other estimation methods, except for the oracle scenario. This pattern continues across all tested values of $\alpha$, indicating greater test power by focusing on the most informative instruments. The improved power in rationality and symmetry testing aligns with the theoretical advantages of P-GMM, which prioritize moment relevance and validity, thus enhancing estimator efficiency and hypothesis testing robustness. Finally, as the sample size increases to $T=2000$, the performance of P-GMM estimation improves further, with standard errors decreasing and the power of the $J$-tests increasing. This trend validates P-GMM's capability to maintain theoretical rigor and practical efficacy in managing instrument selection, optimizing performance in both small and large samples. These findings affirm that P-GMM estimation not only delivers strong finite sample properties in terms of bias and precision but also optimally balances\\ test power and efficiency, even within complex data structures as represented by DGP 2.\looseness=-1

\section{Evaluating Forecast Rationality with SPF Data}\label{sec:empirl}
\noindent The rational expectations hypothesis assumes that agents incorporate all available information when generating forecasts. Macroeconomic forecasts from established sources, particularly those by professional forecasters, play a critical role in shaping economic outlooks and decision-making processes \citep{C03}. This importance underscores the need to empirically evaluate the rationality of forecasts, particularly those from the SPF, given that professional forecasters' incentives and expertise distinguish them in economic modeling. The SPF is particularly well-suited for rationality testing because its participants, unlike other economic agents such as households, are incentivized to reveal accurate beliefs and are typically more informed \citep{KR90}. In this section, we employ the proposed P-GMM estimation within a forecast rationality context, using SPF data for empirical illustration. To streamline the GMM optimization, we utilize the identity matrix as a weighting matrix and limit the forecast horizon to a one-quarter horizon ($h=1$). The analysis draws on data from the Federal Reserve Bank of Philadelphia's SPF, the oldest quarterly survey of macroeconomic forecasts in the United States, which commenced in the fourth quarter of 1968 (denoted as 1968:Q4). Survey participants in the SPF provide point forecasts on various macroeconomic indicators, some of which have been included since the survey's inception, while others were added beginning in the third quarter of 1981. There is a substantial body of literature on the analysis of data from this survey; see \cite{EKT08}, \cite{GW09}, \cite{WL14}, \cite{RS16}, \cite{DHK22}, among others. In our analysis, we assume a linear forecasting model, consistent with the SPF's structured data collection. Additionally, we assume that forecasters operate under a quadratic loss function ($p_0=2$), reflecting the common assumption of symmetry in forecast error preferences. This structure enables a rigorous examination of P-GMM's application within the rationality frame-\\work, allowing for robust assessment of the relevance and validity of selected instruments.\looseness=-1

We focus on business cycle forecasts and the following series are chosen consequently as the target variable $(Y_{t+1})$: the quarterly growth rates for nominal GNP/GDP (1968:Q4-2019:Q4), the quarterly growth rates for the price index for GNP/GDP (1968:Q4-2019:Q4), the quarterly growth rates for consumption (1981:Q3-2019:Q4), and the quarterly growth rates for real residential fixed investment (1981:Q3-2019:Q4). All values are obtained from the most recently revised data vintage (2024:Q3), which
should represent past economic conditions as accurately as possible from the current perspective. We restrict our analysis to data until 2019:Q4. This exclusion prevents distortions from the extreme fluctuations induced by the Covid-19 pandemic, which caused substantial disruptions to economic activity and introduced atypical variations in many macroeconomic indicators. Given that the SPF encompasses predictions from a broad array of professional forecasters, the forecasts inherently exhibit a degree of dispersion across individual predictions; see \cite{CT09}, \cite{PT10}, \cite{WL14}, and \cite{CC24}. In this paper, we adopt the mean response from each survey as the consensus forecast, thereby mitigating individual forecasting biases and providing a collective, representative forecast for each target variable. The growth rates are calculated as the difference between natural logarithms of subsequent values. The periods $t$ and $t+1$ refer, respectively, to the quarters in which forecasts are released and the subsequent quarters to\\ which these forecasts apply.\looseness=-1

\vspace{0.1cm}

{\captionof{table}{SPF Sample Statistics} \label{tab:tab6}
\vspace{-0.1cm}
\begin{table}[h!]
\renewcommand\arraystretch{1.6}
\begin{center}
\resizebox{15.2cm}{!}{
\begin{tabular}{ccccccc p{1cm}}
\hline\hline
   $Y_{t+1} (\%)$                 & Mean  & Median  & Variance &Skewness & Kurtosis   & Min & Max  \\ \hline
GNP/GDP Growth Rate     &1.5465         & 1.4001  & 0.9138 & 0.6880       & 5.9283   & -1.9518 &  5.8471 \\ \hline
Price Index Growth Rate      &0.8531       & 0.6345  & 0.3751   & 1.2996       & 4.3428   & -0.1130 & 3.2000        \\ \hline
Consumption Growth Rate & 0.7373  & 0.7228   & 0.2847       & -0.2356   & 3.8013    & -0.8996 & 2.0368 \\  \hline
Investment Growth Rate & 0.4910  & 0.8658   & 14.5286      & -0.0093   & 5.8241    & -10.4390 &  17.2089     \\  \hline\hline
\end{tabular}}
 \end{center}
 \vspace{-0.8cm}
\end{table}}

\begin{figure}[htp!]
\begin{center}
\subfigure{
\includegraphics[scale=0.26]{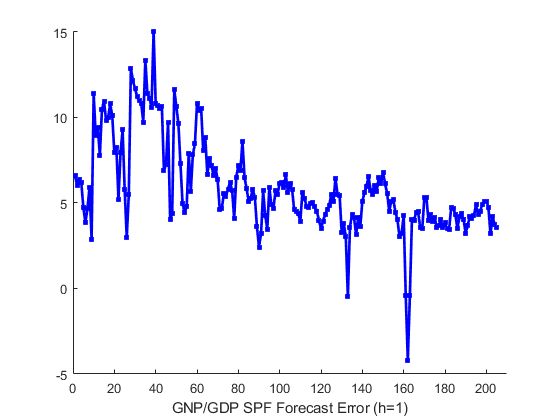}}
\subfigure{
\includegraphics[scale=0.26]{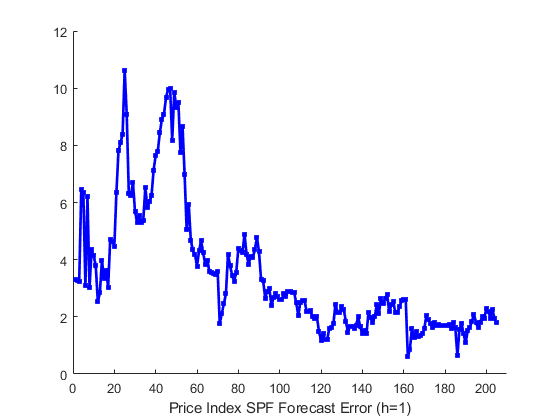}}
\subfigure{
\includegraphics[scale=0.26]{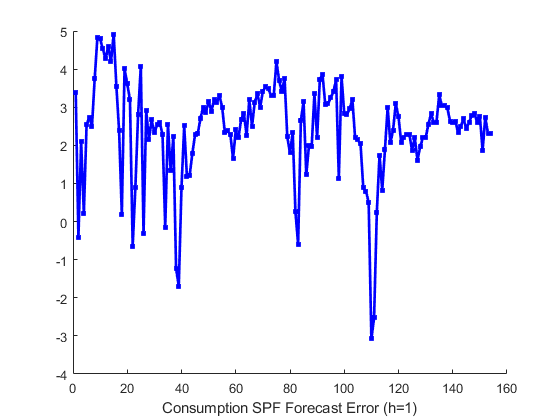}}
\subfigure{
\includegraphics[scale=0.26]{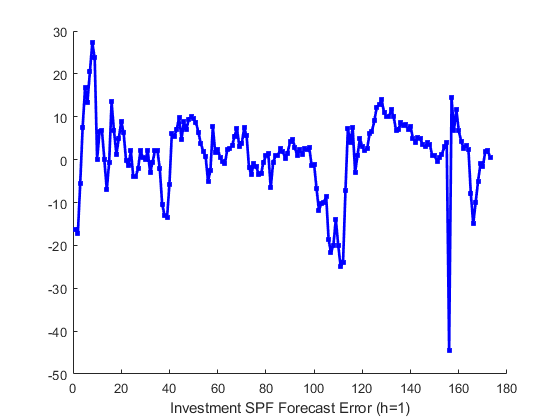}}
\vspace{-0.1cm}
\caption{SPF Forecast Errors for Different Variables}\label{fig:emp1}
\end{center}
\vspace{-0.425cm}
\end{figure}

Table \ref{tab:tab6} summarizes key descriptive statistics for each variable, illustrating notable differences in variance and distributional properties. For instance, the GNP/GDP growth rate displays the highest mean (1.5465) and median (1.4001), reflecting more stable growth compared to the other series. The investment growth rate exhibits the largest variance and kurtosis among the selected series, suggesting that it is inherently more volatile and susceptible to economic fluctuations than other variables. This finding aligns with existing literature on investment cycles, which are typically more sensitive to economic conditions. Before undertaking the estimation and testing of forecast rationality, we conduct a stationarity assessment using the unit root test developed by \cite{ERS96}. Our results indicate that all growth rate series reject the null hypothesis of a unit root, confirming their stationarity. Forecast errors are subsequently calculated as the difference between actual values and one-step-ahead forecasts (Figure \ref{fig:emp1}), where the forecast represents the average prediction across all SPF participants. Figure \ref{fig:emp1} visually captures forecast errors over time, highlighting significant deviations during the 2008 financial crisis. This period of heightened economic uncertainty and unexpected market dynamics led to substantial forecast errors, as forecasters struggled to account for rapidly changing conditions. However, in the quarters immediately following the crisis, forecast errors visibly decrease, suggesting a swift adaptation by forecasters who recalibrated their models to better accommodate the post-crisis economic landscape. Notably, the forecast errors for investment growth are especially volatile, which aligns with the high variance and kurtosis values observed in Table \ref{tab:tab6}. This preliminary statistical and graphical examination underscores the utility of the SPF data for analyzing forecast rationality under the proposed P-GMM framework. By capturing both the dispersion among individual forecasts and the adjustment process following major economic events, our approach leverages the strengths of SPF data to rigorously evaluate forecast accuracy and the relevance of instrumental variables within a stru-\\ctured rationality testing framework.\looseness=-1

In assessing the instruments used for estimation and forecast rationality testing, we construct a high-dimensional instrument set that includes a constant, absolute lagged forecast errors, three lags of forecast errors, lagged changes in the actual values of all macroeconomic variables (real GDP, price index, consumption, and investment), as well as lagged changes in forecasts themselves. To facilitate comparison, we present the results for three distinct scenarios: (i) using all available instruments, denoted as ``All''; (ii) using only a constant and the lagged change in actual values, labeled as ``CON+IV''; and (iii) implementing the proposed P-GMM estimation approach. The resulting estimates of the asymmetry parameter $\alpha$, along with their standard errors and the corresponding rationality test outcomes, are summarized in Table \ref{tab:tab5}, providing a comprehensive view of the impact of instrument sele-\\ction on forecast rationality and symmetry testing.\looseness=-1

\vspace{0.1cm}

{\captionof{table}{Estimation and Testing for Symmetry and Rationality} \label{tab:tab5}
\vspace{-0.1cm}
\begin{table}[h!]
\renewcommand\arraystretch{1.36}
\begin{small}
\begin{center}
\resizebox{13.8cm}{!}{
\begin{tabular}{ccccccc p{5cm}}
\hline\hline
   $Y_{t+1} (\%)$                 & Method  & $\alpha$  & se &$t$-Stat & $J$-Stat   & $p$-value ($J$-Stat)  \\ \hline
GNP/GDP Growth Rate     &All         & 0.4627  & 0.0493 & -0.7566       & 2.1576   & 0.7429 \\
&CON+IV         & 0.4649  & 0.0507 & -0.6923       & 0.3782   & 0.7163      \\
              & P-GMM  & 0.5276   & 0.0421 & 0.6556   & 0.0069   & 0.8452      \\ \hline
Price Index Growth Rate      &All       & 0.5823  & 0.0846   & 0.9728       & 3.9582   & 0.6419       \\
&CON+IV         & 0.5267  & 0.0677 & 0.3944  & 1.5437   & 0.3590     \\
            & P-GMM    &  0.5028   & 0.0481 & 0.0582  & 2.0963    & 0.7724       \\ \hline
Consumption Growth Rate & All  & 0.3662   & 0.0610       & -2.1934   & 7.4782    & 0.0725       \\
&CON+IV         & 0.3305  & 0.0426 & -3.9789 & 2.2563   & 0.0629      \\
& P-GMM & 0.2253  &0.0283   &-9.7067  &4.5980 & 0.0402\\  \hline
Investment Growth Rate & All  & 0.5472   & 0.0814      & 0.5799   & 0.1028    & 0.8254       \\
&CON+IV         & 0.5132  & 0.0769 & 0.1717 & 0.1826   & 0.6402      \\
& P-GMM & 0.3487  &0.0425   &-3.5600  &0.1626 & 0.6710\\  \hline\hline
\end{tabular}}
 \end{center}
 \vspace{-0.08cm}
 \end{small}
{{\bf{\textit{Note}}}: In the table, ``se'' denotes standard error, ``CON'' indicates a constant instrument, ``IV'' represents lagged change in actual values, ``$t$-Stat'' is the $t$-statistic used to assess whether the estimated asymmetry parameter $\alpha$ deviates significantly from 0.5, and ``$J$-Stat'' is the $J$-statistic employed to evaluate forecast rationality. In the method column, ``All'' signifies estimation using all available moment conditions without selection, ``CON+IV'' refers to estimation based on a constant and the lagged instrument, and ``P-GMM'' denotes estimation using the proposed approach.\looseness=-1}
\end{table}}

\vspace{-0.06cm}

To determine whether the asymmetry parameter $\alpha$ significantly deviates from 0.5 and to quantify the uncertainty around this estimate, we conduct a $t$-test for each forecasting scenario, enabling us to evaluate whether any observed deviations are statistically significant or merely reflect sampling variability. For GNP/GDP growth rate, the ``All'' and ``CON+IV'' methods produce estimates of $\alpha$ that are slightly below 0.5, suggesting greater sensitivity to under-predictions; however, these estimates are not significantly different from 0.5, indicating a roughly symmetric cost structure for forecast errors. The P-GMM estimation yields an $\alpha$ value marginally above 0.5, which also supports the hypothesis of symmetry in loss. This consistency across methods implies that for GNP/GDP growth rates, forecasters may assign similar costs to both over- and under-predictions, thus indicating a balanced approach in their cost functions. For price index growth rate forecasts, all three estimation methods yield $t$-statistics that confirm symmetry in the loss function, reinforcing the notion that forecasters view errors on both sides equally costly. This outcome suggests that when predicting price levels, forecasters likely incorporate a symmetric consideration of costs associated with potential over- or under-estimations, reflecting a stable and balanced view on inflation-related measures. In the case of consumption growth rate forecasts, results reveal a markedly different pattern. Across all estimation methods, the asymmetry parameter $\alpha$ deviates considerably from 0.5, clearly indicating an asymmetric loss function that places greater weight on negative forecast errors. This asymmetry highlights a cautious approach in consumption forecasting, where forecasters may avoid overestimating growth due to the potential high economic costs of underestimating consumer demand, which could have adverse effects on business planning and economic stability. For investment growth rate forecasts, the P-GMM results reveal a marked difference from those based on the ``All'' and ``CON+IV'' estimations. While the latter two methods yield values of $\alpha$ close to 0.5, indicating symmetric preferences, the P-GMM approach suggests a distinctly asymmetric loss function, with $\alpha$ significantly below 0.5. This discrepancy highlights the potential for P-GMM to capture nuances in forecast behavior that may be obscured when all instruments are included without selection. In addition, the P-GMM estimate aligns with practical concerns in investment forecasting, i.e., overestimating growth could lead to costly excesses in inventory or capital, while underestimation might lead to missed investment opportunities. This disparity in costs likely causes forecasters to apply differential weights to positive and negative forecast errors, resulting in an asymmetric loss function that P-GMM is uniquely suited to reveal by optimizing the selection of rel-\\evant and valid instruments.\looseness=-1

Furthermore, the reduced standard errors in P-GMM estimation for the considered series underscore the method's efficiency in parameter estimation. The P-GMM approach benefits from its capacity to select moments that satisfy the conditions of validity and relevance (Definitions \ref{def1} and \ref{def2}), effectively reducing the asymptotic variance of the estimator. This technical advantage is consistent with the theoretical properties of P-GMM, where irrelevant instruments---those that do not meaningfully contribute to the identification of the parameter of interest---are systematically excluded, thereby diminishing the potential for overfitting and bias in finite samples. This selective mechanism leads to a more efficient estimation of the asymmetry parameter $\alpha$, with lower standard errors enhancing the precision of $\alpha$ in capturing forecasters' asymmetric loss preferences. By focusing on the most informative instruments, P-GMM not only improves statistical efficiency but also enhances the robustness of the rationality and symmetry tests. This robustness is particularly critical in the context of forecast rationality, as it ensures that the evaluation of forecasters' preferences and biases is reliable across various economic contexts. The P-GMM method thus enables a more granular understanding of forecast behavior by delivering refined estimates that closely reflect forecasters' economic incentives and risk sensitivities, offering a rigorous basis for testing the rationality hypothesis in professional economic forecasting.\looseness=-1

Turning to the forecast rationality test using the $J$-statistic, we observe consistent findings across P-GMM and the alternative methods (``All'' and ``CON+IV'') for GNP/GDP growth rate, price index growth rate, and investment growth rate forecasts. All methods suggest that these forecasts satisfy rational expectations at the 5\% significance level, indicating that forecasters use the available information efficiently when predicting these economic indicators. However, a notable divergence occurs with the consumption growth rate. While the ``All'' and ``CON+IV'' methods do not reject rationality, the P-GMM $J$-test does reject rationality at the 5\% significance level. This discrepancy may stem from the presence of unselected instruments in the other two methods, which could introduce noise and reduce the test's power. In the P-GMM estimation, by contrast, the $p$-value associated with the $J$-statistic is significantly lower, suggesting that the more targeted selection of valid and relevant moments enhances the sensitivity of the rationality test. This efficiency gain aligns with the theoretical expectation that\\ P-GMM estimation improves model performance by excluding uninformative instruments.\looseness=-1

Note that the empirical analysis suggests that SPF data remains valuable for consumption forecasts but underscores the importance of carefully selecting relevant instruments to enhance forecast rationality and precision. The P-GMM approach shows that traditional methods might either include irrelevant instruments or miss crucial information, potentially obscuring true biases in consumption forecasts. Specifically, for consumption, where forecast errors can carry substantial economic costs, the targeted selection of instruments helps capture any asymmetries in forecasters' loss functions, leading to more reliable forecasts. Rather than discouraging the use of SPF data, the findings recommend refining model specifications to leverage the most informative instruments, which can ultimately improve\\ both the accuracy of consumption forecasts and the robustness of rationality assessments.\looseness=-1

\section{Conclusions} \label{sec:conclusion}
\noindent In economics, the foundational assumption that agents form rational expectations has motivated a substantial body of work dedicated to evaluating the accuracy and rationality of economic forecasts. Given that the number of instruments in estimation and testing often exceeds the number of parameters of interest, effectively identifying a subset of instruments that optimally informs forecast rationality remains a key challenge. This paper addresses these issues by incorporating a rigorous moment selection approach within the framework of EKT (2005), specifically employing P-GMM estimation to identify valid and relevant moments without presuming instrument validity. By selecting only the most informative moment conditions, P-GMM estimation reduces biases and inefficiencies associated with over-instrumentation, a common problem in traditional GMM applications. The selection consistency and asymptotic normality are established within the framework of EKT (2005). Monte Carlo simulations validate the efficacy of this approach, demonstrating that P-GMM successfully identifies all relevant moment conditions, enhances estimation precision, and strengthens test power. Furthermore, the empirical analysis underscores the practical value of moment selection in real-world forecast settings, where instrument validity and relevance are often unknown. In this context, the P-GMM approach offers a powerful solution for extracting reliable information from potentially noisy data, leading to more\\ accurate estimators of forecasters' preferences, especially in asymmetric forecast loss functions.\looseness=-1

This paper demonstrates that for external users aiming to assess the rationality of third-party forecasts, such as those in the Greenbook, P-GMM estimation can serve as an effective tool to obtain consistent and efficient results. Although P-GMM proves valuable for selecting valid and relevant instruments, other methodologies like the two-stage moment selection approach by \cite{LSL10} and \cite{GLT20} also merit exploration for instrument selection. A comparative analysis of these methods could yield insights into their respective strengths in accurately identifying forecasters' attitude parameters. Additionally, while our focus here is on time series forecasting, there is potential to extend this work by integrating moment selection with panels of forecasts, as proposed by \cite{TZ19}. This integration would enable rationality testing across both time series and cross-sectional dimensions, offering richer insights into forecast rationality across variables or periods. Finally, it is important to note that our empirical analysis evaluates forecast errors using the most recently revised (final) data, which may differ from the information available to forecasters at the time of prediction. While this approach is common in the literature, it does introduce a discrepancy between the information set available to forecasters at the time of forecasting and the data used to evaluate their accuracy. As such, the rationality test outcomes presented are conditional on final data. Previous studies (e.g., \cite{CS01}) have shown that the use of real-time vintages may yield different conclusions, particularly when early data revisions are substantial. Extending our analysis to incorporate real-time data\\ availability remains an important avenue for future work.\looseness=-1

\section*{{\normalsize{Declaration of Competing Interest}}}
\noindent The authors declare that they have no known competing financial interests or personal relationships that could have appeared to influence the work reported in this paper.\looseness=-1

\section*{\normalsize{Appendix: Technical Proofs}}
\noindent {\bf{Proof of Lemma 1}}.  Let \( \hat{B}_{G+\ell} = [\hat{B}_{G}, g_{\ell}] \in \mathbb{R}^{k_0 \times (m+1)} \), where \( g_{\ell} \) is the new moment vector associated with the additional moment condition. To justify the inequality, we rely on the fact that adding a valid moment condition (i.e., one that contributes positively to the estimator's efficiency shown in Definition \ref{def2}) leads to a positive semi-definite increment in the information matrix \( \hat{B}' \hat{S}^{-1} \hat{B} \). In general, the positive semi-definite matrix \(\hat{S}_{G+\ell}\), which incorporates the weighting adjustments due to the addition of\\ the new moment condition, can be structured as\looseness=-1
\[
\hat{S}_{G+\ell} = \begin{bmatrix} \hat{S}_{G} & s_{G,\ell} \\ s_{G,\ell}' & s_{\ell,\ell} \end{bmatrix},
\]
where $\hat{S}_{G} \in \mathbb{R}^{m \times m}$ is the original weighting matrix for the ``Good'' instruments, $s_{G,\ell} \in \mathbb{R}^{m \times 1}$ represents the cross-term between the ``Good'' instruments and the new moment $\ell$, and $s_{\ell,\ell} \in \mathbb{R}$ is the vari-\\ance of the additional moment. The Schur complement of \(\hat{S}_{G}\) in \(\hat{S}_{G+\ell}\) is given by\looseness=-1
\[
S_{\text{Schur}} = s_{\ell,\ell} - s_{G,\ell}' \hat{S}_{G}^{-1} s_{G,\ell}.
\]
The matrix \(\hat{S}_{G+\ell}\) is positive semi-definite if and only if \(S_{\text{Schur}} \geq 0\). This implies that
\vspace{-0.1cm}
\[
\hat{S}_{G+\ell}^{-1} \leq \begin{bmatrix} \hat{S}_{G}^{-1} & 0 \\ 0 & 0 \end{bmatrix},
\]
which essentially indicates that incorporating the new moment condition (if valid) does not reduce the positive semi-definiteness of \(\hat{S}_{G+\ell}\) compared to \(\hat{S}_{G}\). Now consider the information matrix after adding\\ the moment condition, \( \hat{B}'_{G+\ell} \hat{S}^{-1}_{G+\ell} \hat{B}_{G+\ell} \) and represent it as\looseness=-1
\[
\begin{bmatrix} \hat{B}_{G} & g_{\ell} \end{bmatrix} \begin{bmatrix} \hat{S}_{G}^{-1} + \hat{S}_{G}^{-1} s_{G,\ell} (s_{\ell,\ell} - s_{G,\ell}' \hat{S}_{G}^{-1} s_{G,\ell})^{-1} s_{G,\ell}' \hat{S}_{G}^{-1} & -\hat{S}_{G}^{-1} s_{G,\ell} (s_{\ell,\ell} - s_{G,\ell}' \hat{S}_{G}^{-1} s_{G,\ell})^{-1} \\ -(s_{\ell,\ell} - s_{G,\ell}' \hat{S}_{G}^{-1} s_{G,\ell})^{-1} s_{G,\ell}' \hat{S}_{G}^{-1} & (s_{\ell,\ell} - s_{G,\ell}' \hat{S}_{G}^{-1} s_{G,\ell})^{-1} \end{bmatrix} \begin{bmatrix} \hat{B}_{G} \\ g_{\ell} \end{bmatrix}.
\]
Expanding this product, we find that
\[
\hat{B}'_{G+\ell} \hat{S}^{-1}_{G+\ell} \hat{B}_{G+\ell} = \hat{B}'_{G} \hat{S}^{-1}_{G} \hat{B}_{G} +\underbrace{\hat{B}'_{G} \hat{S}^{-1}_{G}s_{G,\ell}S_{\text{Schur}}^{-1}s'_{G,\ell} \hat{S}^{-1}_{G}\hat{B}_{G}+g'_{\ell}S_{\text{Schur}}^{-1}g_{\ell}}_{\text{nonnegative terms}}.
\]
Since \(S^{-1}_{\text{Schur}} \geq 0\), the additional terms introduced by $g_{\ell}$ and $s_{G,\ell}$ are nonnegative. This indicates that adding a new, valid moment condition indeed increases or maintains the total information, yielding the\\ desired inequality $\hat{B}'_{G+\ell} \hat{S}^{-1}_{G+\ell} \hat{B}_{G+\ell} \geq \hat{B}'_{G} \hat{S}^{-1}_{G} \hat{B}_{G}$. This completes the proof.\looseness=-1

$\hfill\square$

\noindent {\bf{Proof of Theorem 1}}. The proof of Theorem 1 involves analyzing the asymptotic behavior of the penalized objective function in \eqref{eq14} and proving the conditions under which the slackness parameter estimators \(\hat{\beta}_{\ell}\) converge to zero or remain non-zero, depending on the moment condition's classification (valid and relevant, valid but irrelevant, or invalid). Based on these, we consider the three cases for moment conditions as specified by Theorem 1: invalid moments \((\ell \in \mathcal{B}_1)\), irrelevant but valid moments\\ \((\ell \in \mathcal{B}_0)\), and valid and relevant moments \((\ell \in \mathcal{A})\).\looseness=-1

{\bf{Case 1}}: Invalid Moments \((\ell \in \mathcal{B}_1)\). For invalid moments, by Definition \ref{def1}, we know that
\vspace{-0.05cm}
   \[
   \mathbb{E}\left[V_{i,t}\left(\mathbb{I}\left(e_{t+h}^{*}<0\right) - \alpha_0 \right)\lvert e_{t+h}^{*}\rvert^{p_0 - 1}\right] \neq 0.
   \vspace{-0.05cm}
   \]
This condition implies that including an invalid moment introduces bias into the estimator. In the penalized objective function \eqref{eq14}, the effect of this non-zero expectation appears in the penalty term \(\lambda_T \omega_{T,\ell} \lvert \beta_{\ell} \rvert\). Since invalid moments have $\mathbb{E}[\cdot] \neq 0$, as \(T \rightarrow \infty\), the magnitude of \(\dot{\beta}_{T,\ell}\) will tend to be large for invalid moments. Consequently, the penalty term \(\lambda_T \omega_{T,\ell} \lvert \beta_{\ell} \rvert\) for invalid moments will be minimized by setting \(\beta_{\ell} \neq 0\). Here, the small penalty weight \(\omega_{T,\ell}=\dot{\mu}_{T,\ell}^{r_{1}}\lvert \dot{\beta}_{T,\ell}\rvert^{-r_{2}}\) implies a weak incentive for the\\ estimator to drive $\beta_{\ell}$ towards zero. Therefore, as \(T \rightarrow \infty\),\looseness=-1
\vspace{-0.05cm}
   \[
   \operatorname{P}\left(\cup_{\ell \in \mathcal{B}_1}\left\{\hat{\beta}_{\ell} = 0\right\}\right) \rightarrow 0,
   \vspace{-0.05cm}
   \]
meaning the estimator \(\hat{\beta}_{\ell}\) for invalid moments will not converge to zero, and thus these moments will\\ not be selected in the final model.\looseness=-1

{\bf{Case 2}}: Irrelevant but Valid Moments \((\ell \in \mathcal{B}_0)\). For irrelevant but valid moments, by Definition \ref{def2}, we know that\looseness=-1
   \[
   \mathbb{E}\left[(V_{j,t}, V_{i,t})'\left(\mathbb{I}(e_{t+h}^{*}<0) - \alpha_0\right)\lvert e_{t+h}^{*}\rvert^{p_0 - 1}\right] = 0.
   \]
Although these moments are valid, they are irrelevant in terms of improving estimator efficiency. When applying \eqref{eq14}, the penalty coefficient \(\omega_{T,\ell}\) for these irrelevant moments becomes very small due to the definition of \(\dot{\mu}_{T,\ell}\) in \eqref{eq15}, where \(\dot{\mu}_{T,\ell}\) approaches zero for irrelevant moments as the informational content is negligible. Since \(\dot{\mu}_{T,\ell}\) approaches zero for irrelevant moments, the penalty weight \(\omega_{T,\ell}=\dot{\mu}_{T,\ell}^{r_{1}}\lvert \dot{\beta}_{T,\ell}\rvert^{-r_{2}}\) also becomes small. As a result, for irrelevant moments, \(\beta_{\ell}\) is effectively retained\\ with \(\beta_{\ell} \neq 0\) in the optimization problem as \(T \rightarrow \infty\). Thus,\looseness=-1
   \[
   \operatorname{P}\left(\cup_{\ell \in \mathcal{B}_0}\left\{\hat{\beta}_{\ell} = 0\right\}\right) \rightarrow 0 \ \text{as} \ T \rightarrow \infty.
   \]
This ensures that irrelevant but valid moments remain in the model, given that their non-zero status does not interfere with estimation efficiency.\looseness=-1

{\bf{Case 3}}: Valid and Relevant Moments \((\ell \in \mathcal{A})\). For valid and relevant moments, we have both validity (zero expectation) and relevance (significant information contribution). By Definitions \ref{def1} and \ref{def2},\\ valid and relevant moments satisfy\looseness=-1
   \[
   \mathbb{E}\left[(V_{j,t}, V_{i,t})'\left(\mathbb{I}(e_{t+h}^{*}<0) - \alpha_0\right)\lvert e_{t+h}^{*}\rvert^{p_0 - 1}\right] = 0,
   \]
and they contribute significantly to the estimator's efficiency. The relevance criterion in \eqref{eq15} indicates that \(\dot{\mu}_{T,\ell}\) will be large for relevant moments. With a large \(\dot{\mu}_{T,\ell}\), the penalty weight \(\omega_{T,\ell}=\dot{\mu}_{T,\ell}^{r_{1}}\lvert \dot{\beta}_{T,\ell}\rvert^{-r_{2}}\) is substantial. Consequently, the penalty term \(\lambda_T \omega_{T,\ell} \lvert \beta_{\ell} \rvert\) in the objective function is large, and the estimator drives \(\hat{\beta}_{\ell}\) towards zero for all valid and relevant moments as \(T \rightarrow \infty\). As a result, the probabi-\\lity of \(\hat{\beta}_{\ell}\) converging to zero is maximized for valid and relevant moments, meaning\looseness=-1
   \[
   \operatorname{P}\left(\cap_{\ell \in \mathcal{A}}\left\{\hat{\beta}_{\ell} = 0\right\}\right) \rightarrow 1 \ \text{as} \ T \rightarrow \infty.
   \]
This outcome indicates that only valid and relevant moments are retained in the final model.

The selective behavior of the penalty structure in P-GMM ensures that only valid and relevant moments are asymptotically retained. This establishes moment selection consistency for the P-GMM estimator, thereby achieving the goal of systematically incorporating only the moments that optimize esti-\\mation accuracy and efficiency.\looseness=-1

$\hfill\square$

\noindent {\bf{Proof of Theorem 2}}. We proceed by examining the penalized objective function and establishing conditions under which only valid and relevant moments contribute to the asymptotic distribution. Given the penalized objective function in \eqref{eq14}, we have\looseness=-1
\vspace{-0.1cm}
\[
\hat{\alpha}_0 = \arg \min_{\alpha} \Big[ \overline{g}_T(\alpha, \beta)^{\prime} S_T^{-1} \overline{g}_T(\alpha, \beta) + \lambda_T \sum_{\ell \in \mathcal{A} \cup \mathcal{B}_1 \cup \mathcal{B}_0} \omega_{T, \ell} \lvert \beta_{\ell} \rvert \Big].
\]
Using Theorem \ref{thm1} on consistent moment selection, we know (i) \(\beta_{\ell} = 0\) for all \(\ell \in \mathcal{A}\), ensuring that only valid and relevant moments remain in the minimized objective function as $T \rightarrow \infty$; and (ii) for moments in \(\mathcal{B}_1\) (invalid) or \(\mathcal{B}_0\) (irrelevant), the penalty term $\lambda_T\omega_{T, \ell} \lvert \beta_{\ell} \rvert$ increases with \(\lvert \beta_{\ell}\rvert\), driving \(\hat{\beta}_{\ell}\) away from zero and excluding them from the estimator. This exclusion of uninformative moments ensures\\ that only moments contributing to the efficient estimation of \(\alpha_0\) remain.\looseness=-1

Using the remaining valid and relevant moments in \(\mathcal{A}\), the penalized objective function now asym-\\ptotically resembles an oracle GMM objective based on only those moments. Define\looseness=-1
\[
\overline{g}_{\mathcal{A}, T}(\alpha) = \frac{1}{T} \sum_{t=1}^T g_{\mathcal{A}}(V_t, \alpha),
\]
where \(g_{\mathcal{A}}(V_t, \alpha)\) is the moment condition vector restricted to moments in \(\mathcal{A}\). To derive the asymptotic distribution of \(\hat{\alpha}_0\), we examine the behavior of \(\overline{g}_{\mathcal{A}, T}(\alpha)\) around \(\alpha_0\), the true parameter value. By the\\ central limit theorem for GMM estimators, we have\looseness=-1
\[
\sqrt{T} \overline{g}_{\mathcal{A}, T}(\alpha_0) \stackrel{d}{\rightarrow} \mathcal{N}(0, S_{\mathcal{A}}),
\]
where \(S_{\mathcal{A}} = \lim_{T \to \infty} \operatorname{Var}(\sqrt{T} \overline{g}_{\mathcal{A}, T}(\alpha_0))\) represents the asymptotic variance of the moments in \(\mathcal{A}\). Since the objective function now asymptotically depends only on moments in \(\mathcal{A}\), \(\hat{\alpha}_0\) is a consistent estimator of \(\alpha_0\) when restricted to valid and relevant moments. Given \(\overline{g}_{\mathcal{A}, T}(\alpha_0) \approx 0\) as \(T \to \infty\), we obtain\looseness=-1
\[
\sqrt{T} \overline{g}_{\mathcal{A}, T}(\hat{\alpha}_0) = \sqrt{T} (\overline{g}_{\mathcal{A}, T}(\alpha_0) + B_{\mathcal{A}} (\hat{\alpha}_0 - \alpha_0)) \approx 0,
\]
where \(B_{\mathcal{A}} = \mathbb{E} ( \frac{\partial \overline{g}_{\mathcal{A}, T}(\alpha)}{\partial \alpha} \lvert_{\alpha = \alpha_0} )\) is the Jacobian matrix evaluated at $\alpha_0$. Solving for \(\sqrt{T}(\hat{\alpha}_0 - \alpha_0)\), we arrive at\looseness=-1
\[
\sqrt{T}(\hat{\alpha}_0 - \alpha_0) \approx - (B_{\mathcal{A}}^{\prime} S_{\mathcal{A}}^{-1} B_{\mathcal{A}})^{-1} B_{\mathcal{A}}^{\prime} S_{\mathcal{A}}^{-1} \sqrt{T} \overline{g}_{\mathcal{A}, T}(\alpha_0).
\]
As $T \rightarrow \infty$, the distribution of $\sqrt{T}(\hat{\alpha}_0 - \alpha_0)$ converges in distribution to a normal distribution
\[
\sqrt{T}(\hat{\alpha}_0 - \alpha_0) \stackrel{d}{\rightarrow} \mathcal{N}\left(0, (B_{\mathcal{A}}^{\prime} S_{\mathcal{A}}^{-1} B_{\mathcal{A}})^{-1}\right),
\]
where the asymptotic variance \((B_{\mathcal{A}}^{\prime} S_{\mathcal{A}}^{-1} B_{\mathcal{A}})^{-1}\) is derived from the valid and relevant moments in ${\mathcal{A}}$.\\ This completes the proof of Theorem \ref{thm2}.\looseness=-1

\(\hfill\square\)

\end{document}